\documentclass[a4paper,11pt]{article}

\usepackage{jheppub,amsmath, amssymb,amsthm,dsfont,graphicx,mathtools,multirow,rotating,placeins,verbatim} 

\usepackage[normalem]{ulem}

\usepackage{tikz}
\usetikzlibrary{arrows.meta, positioning}

\usepackage{hyperref}
\usepackage{comment}
\usepackage{color}

\usepackage{setspace}
\usepackage{cancel}
\usepackage{graphicx}
\usepackage{enumerate} 

\tikzset{myarrow/.style={->, shorten >=5pt, shorten <=5pt, >=Latex,thick}}

\def\be{\begin{equation}}
\def\ee{\end{equation}}
\def\ba{\begin{eqnarray}}
\def\ea{\end{eqnarray}}
\def\bal{\begin{equation} \begin{aligned}}
\def\eal{\end{aligned}\end{equation} }

\def\bi{\begin{itemize}}
\def\ei{\end{itemize}}

\def\zb{\bar{z}}

\def\w{\omega}

\def\nh{\hat{n}}

\def\ph{\hat{p}}
\def\I{\mathcal{I}}

\def\w{\omega}

\newcommand{\ov}[2]{\overset{\scriptscriptstyle #1}{#2}\vphantom{#1}}

\def\t{\tau}

\def\T{\mathcal{T}}
\def\harm{\text{harm}}

\def\G{\mathcal{G}}

\def\drag{\text{drag}}
\def\scri{\mathcal{I}}

\def\massless{\text{massless}}
\def\grav{\text{grav}}
\def\htilde{\tilde{h}}
\def\div{\text{div}}
\def\rad{\text{rad}}
\def\E{\mathcal{E}}
\def\inn{\text{in}}
\def\out{\text{out}}
\def\M{\mathcal{M}}

\def\nb{\bar{n}}

\def\N{\mathcal{N}}
\def\Nt{\tilde{\mathcal{N}}}

\def\H{\mathcal{H}}
\def\xb{\text{x}}
\def\bondi{\text{Bondi}}
\def\Mzero{\ov{0}{\M}}
\def\Czero{\ov{0}{C}}
\def\sigmazero{\ov{0}{\sigma}}
\def\izero{\ov{0}{i}}

\def\A{\mathcal{A}}

\def\O{\Omega}

\title{An asymptotic proof of the classical log soft graviton theorem}

\author[a]{Gianni Boschetti}
\author[b]{Miguel Campiglia}

\affiliation[a]{Instituto de F\'isica, Facultad  de  Ingenier\'ia, Universidad  de  la  Rep\'ublica, 
Julio Herrera y Reissig 565,  Montevideo,  Uruguay}

\affiliation[b]{Instituto de F\'isica, Facultad  de  Ciencias, Universidad  de  la  Rep\'ublica ,
Igua  4225,  Montevideo,  Uruguay}

\emailAdd{gboschetti@fing.edu.uy}
\emailAdd{miguel.campiglia@fcien.edu.uy}

\abstract{We  present a derivation of the classical log soft graviton theorem  within the asymptotic framework  of Compère, Gralla, and Wei. The proof relies solely on Einstein equations near timelike, spatial, and null infinity, together with matching properties across these regions. The approach is fully covariant under time reversal and  incorporates contributions from incoming soft radiation.  In the absence of incoming memory one recovers the standard  log soft factor,  which features an  asymmetry  between future and past  hard components. From an asymptotic perspective, the origin of this asymmetry lies in a long-known discontinuity of the gravitational field at spatial infinity.}

\begin{document}
\maketitle
\flushbottom

\section{Introduction}

\begin{figure}[ht]
  \centering
  \includegraphics[width=0.8\textwidth]{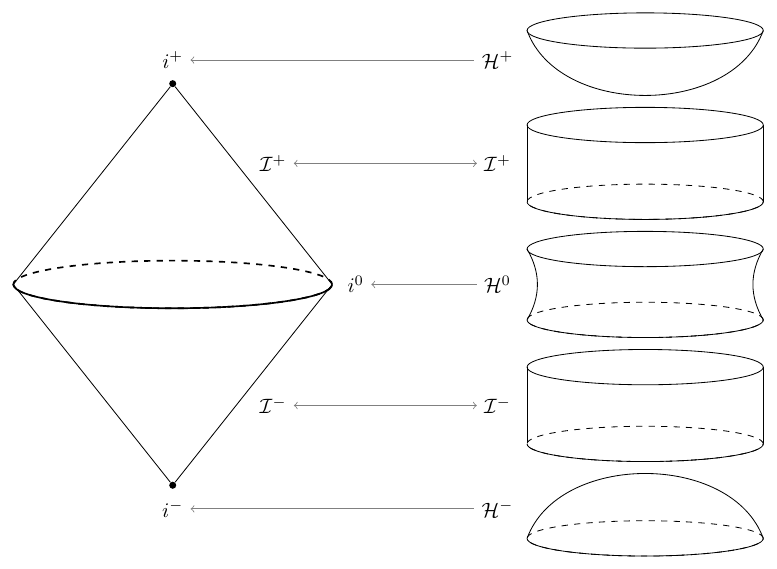}
  \caption{Left: Penrose diagram of an asymptotically flat spacetime. Right: A ``democratic" representations of  null,  timelike and spatial  infinities.  In both cases the null infinities $\I^\pm$ are described by cylinders. 
The  timelike $\H^\pm$ and spatial  $\H^0$ infinities on the right  diagram are hyperboloids   parametrizing the  directions of asymptotic geodesics  at  the points  $i^\pm$ and $i^0$ on the left diagram.}
  \label{figure1}
\end{figure}

In \cite{penrose} Penrose introduced his celebrated conformal description of spacetime, partly motivated by “a longer term aim \ldots for a covariant S-matrix theory incorporating gravitation”. This marked the beginning of what may be called an \emph{asymptotic} approach to gravitational scattering,  which avoids relying on  the Minkowski background built into perturbative treatments. Traditionally, the successes of this approach were mostly kinematical, e.g.  the characterization of  asymptotic states and charges  \cite{aaprl,AS,aajmp}. 
Over the past decade it has become clear that the asymptotic perspective is specially suited for addressing dynamical questions within the soft sector of gravitational radiation \cite{stromgravscatt}. In particular, a variety of \emph{soft theorems} admit a natural formulation in terms of matching properties of the gravitational field across timelike, null and spatial infinity \cite{stromlectures}. In this paper we deepen this viewpoint by presenting an asymptotic proof of the classical log soft graviton theorem \cite{laddhasen1,sahoosen,proofdeq4,senreview}.\footnote{See \cite{logwcl,sayalicons,sayaliqedgrav,fuentelogscalar,comperelogem,bricenolog,Duary:2025siq} for  related discussions in the context of  electromagnetism and massless scalars.} We build on and further develop the framework put forward by Compère, Gralla and Wei (CGW) \cite{cgw} where the three types of infinities are treated on equal footing (see  Figure~\ref{figure1}\footnote{The diagrams  in Figure~\ref{figure1}  only intend to capture the exterior  boundaries of spacetime, i.e. those associated to geodesics that escape to infinity. There can also be internal boundaries, notably in presence of black holes.  In the asymptotic picture, black holes (or any other  compact object) are registered as point-like sources at $\H^\pm$ with multipolar moments appearing at subleading order in the large time expansion. We thank G. Compère for discussions on this point.}). We refer to  \cite{hansen,persides} for earlier work in this direction and \cite{compererobert} for a very recent treatment addressing similar aspects to those discussed here.

Although our contribution is limited to the soft sector, we are motivated by the hope that advancing the CGW framework may  serve the longer-term aim of a fully covariant theory of gravitational scattering. 
This perspective is in line with recent efforts towards boundary  descriptions of the S-matrix   \cite{Kim:2023qbl,Jain:2023fxc,Kraus:2024gso,Ammon:2025avo} in the context of flat space holography (see  \cite{raju,Donnay:2023mrd,Bagchi:2023cen} and references therein).

The plan of the paper is as follows. Section \ref{prelsec} provides some basic preliminaries that permeate our discussion, including notation and conventions. In section \ref{softthmsec} we review the leading and log classical soft graviton theorems, extending the latter to the case where there is incoming soft gravitational  radiation. Section \ref{stimeasymssec} is a self-contained exposition of the main points of the CGW framework that we shall need for the asymptotic proof, along with several extensions, notably explicit Green's functions solutions to the asymptotic metric coefficients at timelike and spatial infinities.  The results of this analysis are put together in section \ref{asymproofsec} to provide a proof of the log soft theorem. 
The paper is supplemented with six appendices containing  technical material.

\section{Preliminaries} \label{prelsec}

\subsection{Asymptotic geodesics}

 In asymptotic Cartesian coordinates, the general asymptotic form of geodesics that escape to infinity is
\be \label{gralasymgeod}
X^\mu(s) \stackrel{|s| \to \infty}{=} s V^{\mu} + \log  |s|  \, c^\mu(V) + O(s^0),
\ee
where $s$ is an affine parameter and  $V^{\mu}$ the asymptotic direction.  $c^\mu$  captures the leading deviation from a straight line,  which we refer to as  \emph{log deviation vector}.\footnote{In terms of asymptotic Christoffel symbols,  $c^\mu(V) = \lim_{|s| \to \infty} s^2 \Gamma^\mu_{\nu \rho}(s V) V^\nu V^\rho$. Our minimal assumption on asymptotic flatness is that this limit is well defined.} This vector is intimately tied with \emph{logarithmic translations} \cite{Bergmann:1961zz,aalog}, diffeomorphisms that have the asymptotic form
\be \label{deflogvf}
\xi^\mu_L \stackrel{|s| \to \infty}{=} \log |s|\, L^\mu + \cdots, 
\ee
and that act on $c^\mu$ according to 
\be \label{delLcmu}
\delta_L c^\mu = -L^\mu.
\ee

The possible asymptotic directions $V^\mu$ fall into five categories: future/past null, future/past timelike and spacelike, corresponding to the five asymptotic boundaries pictured in the right panel of Figure \ref{figure1}.  

For null rays,  we parametrize the asymptotic  directions by
\be \label{defnmu}
n^\mu = (1,\nh),
\ee
where $\nh$ is a unit 3-vector. Eq.  \eqref{gralasymgeod} then takes the form
\be \label{nullraygral}
X^\mu(s)  \stackrel{s \to \pm \infty}{=} s \, n^\mu  +  \log  |s|  \, c^\mu_{\scri^\pm} + O(s^0) \quad \text{(null geodesics)},
\ee
where incoming/outgoing null rays are distinguished by the sign of $s$. A key property of asymptotic null geodesics  is that their log deviation vector $ c^\mu_{\scri^\pm}$ is  independent of the asymptotic direction \cite{sahoosen,gianni1}.  This allows one to  ``gauge it away'' by a log translation, as expected by general results on the existence of  radiative coordinates \cite{blanchetnull}.  In such (future/past) \emph{radiative log frame}, the $O(s^0)$ piece of \eqref{nullraygral} can  be used to distinguish different arrival/departure times. By an appropriate choice of $O(s^0)$ parametrization, Eq. \eqref{nullraygral}  can be brought into the form
\be \label{Xsnullradframe}
X^\mu(s)  \stackrel{s \to \pm \infty}{=} s \, n^\mu  + u \, t^\mu + \cdots \quad \text{(null geodesics in  radiative log frame)},
\ee
where $t^\mu = (1,\vec{0})$,  $u$ is a retarded/advanced time  and  the dots refer to terms that vanish in the large $s$ limit.  The ``endpoints'' of such null geodesics, parametrized by $(u,\nh)$, define the null infinities $\scri^\pm$.

For  non-null geodesics, we normalize the asymptotic velocity such that $s$ becomes a proper time/distance.  The log deviation vector now exhibits a non-trivial dependance on the asymptotic velocity  and cannot  be gauged away.\footnote{This velocity-dependence is related to   the non-smoothness of the Penrose metric at the points $i^0,i^\pm$ \cite{hansen}.} Eq.  \eqref{gralasymgeod} then  takes the form
\ba 
X^\mu(s) & \stackrel{s \to \pm \infty}{=} & s V^{\mu}  +  \log  |s| c^\mu_{\H^\pm}(V) + O(s^0),  \quad V^\mu V_\mu = -1, \quad V^0>0, \label{Xstimelike} \\
& &\nonumber \\
X^\mu(s)  & \stackrel{s \to \infty}{=}  & s V^{\mu}  + \log  |s| c^\mu_{\H^0}(V) + O(s^0),  \quad  V^\mu V_\mu = 1 , \label{Xsspacelike}
\ea
for timelike/spacelike geodesics.  Their corresponding ``endpoints''  are parametrized by the unit hyperboloids  $\H^\pm$  and $\H^0$.  Note that, as in the null case, our conventions are such that $V^0>0$ for both incoming and outgoing timelike geodesics. \\

In the following subsection we present the coordinates that we shall use on these ``endpoints at infinity'' spaces,  see  Table \ref{table0} for a summary. The extension of these coordinates into spacetime is discussed in section \ref{stimeasymssec}.

\begin{table}[h!]
\centering
\begin{tabular}{c|l|l}
 space & coordinates & infinity type \\
\hline
\hline
  $\I^\pm$ & $(u,\phi^A)$& null\\
\hline 
 $\H^\pm$ & $(\rho,\phi^A)=x^a$& timelike \\
\hline
  $\H^0$ & $(\t,\phi^A)= x^a$ & spatial
\end{tabular}
\caption{Coordinates  at the five infinities.}
\label{table0}
\end{table}

\subsection{Coordinate conventions}

All five infinities feature an angular direction. We denote by \( \phi^A \)   the corresponding  2d coordinates. These specify the unit  3-vector \( \nh \) and  associated null direction $n^\mu$ \eqref{defnmu}.  The null infinities are therefore parametrized by $(u,\phi^A)$ where $u$ is a retarded/advanced time. For the most part the specific choice of $\phi^A$ will be irrelevant. However, some expressions simplify considerably in holomorphic coordinates $(z,\zb)$ where
\be \label{nhitozzb}
\nh = \frac{1}{1+|z|^2} \left(  z+ \zb, -i (z- \zb), 1- |z|^2 \right) .
\ee

For the unit timelike hyperboloids we define 3d coordinates  $x^a=(\rho,\phi^A)$ by
\be \label{Vxtime}
V^\mu(x) = \big(\sqrt{\rho^2+1}, \rho \, \nh(\phi) \big) \quad  \iff  \quad V^\mu V_\mu = -1, \quad V^0>0,
\ee
where $\rho \in [0,\infty)$ is a ``boost radius''. 

For the unit spatial hyperboloid we take $x^a=(\t,\phi^A)$ with
\be \label{Vxspatial}
 V^\mu(x) = \big(\t, \sqrt{\t^2+1} \, \nh(\phi) \big) \quad  \iff  \quad  V^\mu V_\mu = 1,  \phantom{, \quad V^0>0,}
\ee
where $\t \in (-\infty,\infty)$ is a ``boost time''.

A key element of our discussion will be to understand the behavior of fields on these spaces as one approaches their asymptotic boundaries. At timelike infinity this boundary is denoted as $\partial \H^\pm$ and reached when $\rho \to \infty$. In this limit  the direction \eqref{Vxtime} asymptotes to,
\be \label{limrhoinfVmu}
V^\mu(\rho,\phi) \stackrel{\rho \to \infty}{=} \rho \, n^\mu(\phi) + O(1/\rho).
\ee

At spatial infinity there are two boundaries denoted by  $\partial_\pm \H^0$,  corresponding to $\t \to \pm \infty$.  In these limits, the  direction  \eqref{Vxspatial} becomes
\ba
V^\mu(\t,\phi) &\stackrel{\t \to + \infty}{=} &\t \, n^\mu(\phi) + O(1/\t),  \label{limtaupinfVmu} \\
V^\mu(\t,\phi) &\stackrel{\t \to - \infty}{=} &\t \, \A_* n^\mu(\phi) + O(1/\t) \label{limtauminfVmu} 
\ea
where $\A$ is the antipodal map on the the sphere and $\A_*$ its corresponding pullback,
\be \label{defantipodalmap}
\A_* \nh(\phi) = \nh(\A\phi) = -\nh(\phi).
\ee

It is interesting to see how the null geodesics \eqref{nullraygral} are recovered from  the timelike/spatial ones under the above limits. The divergent factors multiplying the null directions in  \eqref{limrhoinfVmu}, \eqref{limtaupinfVmu} and \eqref{limtauminfVmu}  indicate the  need to simultaneously rescale the affine parameter $s$ in \eqref{Xstimelike}, \eqref{Xsspacelike}. Such rescaling, however, does not affect the corresponding log deviation vectors, from which one concludes that 
\bal
\lim_{\rho \to \infty} c^\mu_{\H^\pm}(\rho,\phi) &= c^\mu_{\scri^\pm},  \\
\lim_{\t \to  \pm \infty} c^\mu_{\H^0}(\t,\phi) & = c^\mu_{\scri^\pm} .
\eal

These relations provide a first example of a matching property across infinities.  In this case, the matching can be used to define a global notion of log translations that simultaneously act on all five infinities. In particular, it  allows to discuss log translation invariance  in a  scattering setting, see   \cite{gianni1}  and subsection \ref{softthmsec}.

\subsection{Log translation frames} \label{logframessec}

In actual computations,  log translations are usually gauge-fixed.  A simple way to do so is by prescribing the value of  \emph{either} $c^\mu_{\scri^+}$ or $c^\mu_{\scri^-}$. There is an obstruction to assign independent values to these quantities, due to a well-known discontinuity at spatial infinity  \cite{hansen,tn,cgw} that fixes their  difference by 
\be \label{obstructioncmuspi}
c^\mu_{\scri^+} - c^\mu_{\scri^-} = 4 G P^\mu,
\ee
where $P^\mu$ is the total spacetime momentum, see \cite{gianni1} and section  \ref{spisec} for further details.

We shall deal with three   fixings or \emph{log  translation frames} that are summarized in Table \ref{table1}.  The (future and past) radiative frames were already mentioned in Eq. \eqref{Xsnullradframe}. The harmonic frame is the one  associated to harmonic coordinates and leads to  a  parity-even gravitational potential at spatial infinity \cite{aalog}.

\begin{table}[h!]
\centering
\begin{tabular}{l||c|c}
 Log  frame & \rule[-1.2ex]{0pt}{2.5ex}  $c^\mu_{\scri^+}$ & $c^\mu_{\scri^-}$ \\
\hline \hline
Future radiative  & $0$ & $- 4 G P^\mu $ \\
\hline
Past radiative  &  $ 4 G P^\mu $ & $0$\\
\hline
Harmonic &  $2 G P^\mu $&  $- 2 G P^\mu $ \\
\hline
\end{tabular}
\caption{Values of $c^\mu_{\scri^\pm}$ on three different log translation frames.}
\label{table1}
\end{table}

Looking down either  column in Table \ref{table1} we can identify the log translation \eqref{delLcmu} that interpolates between  frames. Denoting by $c^{\rad \pm}_\mu, c^\harm_\mu$ the corresponding log deviation vectors, we have: 
\ba
 c^{\rad + }_{\mu} - c^{\rad -}_{\mu} &= & - 4 G P_\mu, \label{cradpluscradminus} \\
c^{\harm}_\mu  - c^{\rad \pm}_\mu &= & \pm 2 G P_\mu. \label{charmcradpm}
\ea

\subsection{Asymptotic particles} \label{asympartsec}

Following standard practice  \cite{senreview}, we treat as \emph{particles} the elementary constituents in a scattering processes, even though they can refer to macroscopic bodies or radiation. This perspective is specially suited for the discussion of soft theorems, as they are  insensitive to the internal structure of the scattering entities. 

We  use an index $i$ to label all such ``particles'', including  incoming,  outgoing,  massive and massless. The momentum of a particle with mass $m_i$ and asymptotic trajectory   \eqref{Xstimelike} is defined as\footnote{We will also  use the notation ``$i \in +$'' and ``$i \in -$'' for outgoing/incoming particles respectively.}
\bal\label{defpinout}
p^\mu_{i} &=  m_i V^{\mu}_{i}  & \text{if} \quad i \in \out, \\
 p^\mu_{i} &=  -m_i V^{\mu}_{i} &\text{if} \quad i \in \inn,
\eal
where, following amplitudes conventions,   incoming momenta carry an overall negative sign.  Massless particles can be described by sending $m_i \to 0$ and $\rho \to \infty$ with $E_i = m_i \rho$ held fixed.  The total  momentum is   given by
\be
P^\mu = \sum_{i \in \out} p^\mu_i = -\sum_{i \in \inn} p^\mu_i .
\ee

We denote by $c_i$ the log deviation vector of the $i$-th particle, omitting the labels $\H^\pm$ or $\I^\pm$ used in Eqs. \eqref{nullraygral} and \eqref{Xstimelike}. A well known consequence of this vector is that it leads to a divergence in the  particle's angular momentum \cite{laddhasen1},
 \begin{equation} \label{Jmunudiv}
J_{\mu\nu}^i    \stackrel{s \to \pm \infty}{=}   \log | s | J^{\div \, i}_{\mu \nu} + O(s^0), 
\end{equation}
where
\be \label{defJdiv}
J^{\div \, i}_{\mu \nu} =   c^i_{[\mu} \, p^i_{\nu]} =c^i_\mu p^i_\nu - c^i_\nu p^i_\mu .
\ee
Important for our analysis is the fact that \eqref{defJdiv} is sensitive to log translations. We denote by $J^{\rad \pm \, i}_{ \mu \nu}, J^{\harm \, i}_{\mu \nu}$ the value of this quantity in the corresponding log  frame. According to  Eqs. \eqref{cradpluscradminus} and \eqref{charmcradpm} these are related by
\ba 
J^{\rad + \, i}_{\mu \nu} &=& J^{\rad - \, i}_{\mu \nu} -    4 G     P_{[\mu} \, p^i_{\nu]} ,   \label{JminitoJplus} \\
J^{\harm \, i}_{\mu \nu} &=&  J^{\rad \pm \, i}_{\mu \nu} \pm 2 G P_{[\mu} \, p^i_{\nu]}.  \label{JharmitoJrad}
\ea

\section{Classical soft theorems} \label{softthmsec}
Gravitational radiation is encoded in the leading deviation from the flat metric along null rays. In asymptotic radiative  coordinates of the type \eqref{Xsnullradframe}, the outgoing gravitational waveform is defined by
\be \label{defhout}
h^\out_{\mu \nu}(u,\nh) := \lim_{r \to \infty} r (  g_{\mu \nu}-\eta_{\mu \nu}),
\ee
where $r = s$ is the asymptotic radial distance.   The classical soft theorems refer to universal components in the low frequency expansion of the Fourier transform of \eqref{defhout},
\begin{equation} \label{hmunuomega}
\tilde{h}^\out_{\mu\nu}(\omega, \hat{n}) \stackrel{\w \to 0}{=}   \omega^{-1} \htilde^{(0)}_{\mu\nu}(\hat{n}) + \log \omega \, \htilde^{(\log)}_{\mu\nu}(\hat{n}) + \cdots 
\end{equation}

The leading   term  is described  by Weinberg's soft  theorem \cite{weinberg}
\be \label{hminuone}
\htilde^{(0)}_{\mu\nu}= -  4 G i \sum_i \frac{p^i_\mu p^i_\nu}{p_i \cdot n},
\ee
 and captures what is known as the gravitational memory effect  \cite{thorne,zhibo}.     The logarithmic term in \eqref{hmunuomega}   can be written as a sum of two contributions \cite{laddhasen1,sahoosen},
 \be \label{hlogdevdrag}
\htilde^{(\log)}_{\mu\nu} = \htilde^{(\div)}_{\mu\nu} + \htilde^{(\drag)}_{\mu\nu}.
\ee
 The first one  originates from the divergence in the particle angular momenta  \eqref{Jmunudiv} and reads\footnote{This is of the same form as the  $O(G)$ subleading soft factor \cite{stromingercachazo}, except that it features the divergent angular momentum.}
 \be \label{hdev}
 \htilde^{(\div)}_{\mu\nu} =  -4G \sum_i \frac{p^i_{(\mu}  J^{\div \, i}_{\nu)\rho} n^{\rho}}{p_i \cdot n}.
\ee
 The second contribution arises from the propagation of the   soft radiation on the curved background due to the ``hard'' particles and is given by
 \cite{sahoosen}
\be \label{hdrag}
\htilde^{(\drag)}_{\mu\nu} = - i \, n \cdot c_{\scri^+}   \, \htilde^{(0)}_{\mu\nu}.
\ee

The analysis of  \cite{laddhasen1,sahoosen} derive  these results  through perturbative calculations in  harmonic gauge, where $J^{\div \, i}_{\mu \nu}$ and $c^\mu_{\scri^+}$ are  evaluated in the harmonic log frame. It was however noted in  \cite{gianni1} that  the sum of \eqref{hdev} and \eqref{hdrag} is invariant under log translations, and hence it holds in \emph{any} log frame. In particular, in the future radiative frame where $J^{\div}_{\mu \nu} =J^{\rad +}_{\mu \nu}$ and $c^\mu_{\scri^+}=0$ one has
\be \label{hlnJtilde}
\htilde^{(\log)}_{\mu\nu} = -4G \sum_i \frac{p^i_{(\mu}  J^{\rad + \, i}_{\nu)\rho} n^{\rho}}{p_i \cdot n} .
\ee
 This rewriting makes explicit the  independence of the log soft factor on outgoing massless particles \cite{sahoosen,rewritten} since $J^{\rad +}_{\mu \nu}=0$ for them.

For the asymptotic proof of the log soft theorem we will use yet   a different  rewriting of $\htilde^{(\log)}_{\mu\nu}$.  In the asymptotic analysis of Einstein equations one works with future/past radiative frames for future/past timelike and null infinities. On the other hand,  \eqref{hdev} and  \eqref{hlnJtilde}  express future and past contributions in the \emph{same} log frame.   Using \eqref{JminitoJplus}  on the incoming terms in \eqref{hlnJtilde} allows to express the log soft factor as
\begin{multline} \label{hlnasymfield}
\htilde^{(\log)}_{\mu\nu} = -4G \sum_{\substack{i  \in \out \\ m_i \neq 0}} \frac{p^i_{(\mu}  J^{\rad + \, i}_{\nu)\rho} n^{\rho}}{p_i \cdot n} - 4G \sum_{\substack{i  \in \inn \\ m_i \neq 0}} \frac{p^i_{(\mu}  J^{\rad - \, i}_{\nu)\rho} n^{\rho}}{p_i \cdot n}   \\ - 16 G^2 \left( P \cdot n \sum_{i  \in \inn }    \frac{p^i_{\mu}  p^{i}_{\nu} }{p_i \cdot n} +  P_{\mu}P_{\nu} \right),
\end{multline}
where the sums in the first line are restricted to massive particles since  $J^{\rad \pm \, i}_{\mu \nu}=0$ for outgoing/incoming massless particles.

This is the form of the log soft theorem that naturally emerges in the asymptotic analysis (modulo incoming soft contributions discussed below).  
In that context, the terms in the first line  of \eqref{hlnasymfield} arise from future/past contributions in the corresponding radiative frames, whereas the second line compensates for the mismatch  \eqref{obstructioncmuspi} between future and past frames.

\subsection*{Incoming soft radiation}
Incoming gravitational radiation is encoded in the past infinity version of \eqref{defhout}, which we write as
\be \label{defhin}
h^\inn_{\mu \nu}(u,\nh) := \lim_{r \to- \infty} r (  g_{\mu \nu}-\eta_{\mu \nu}),
\ee
where   $r=s$ is \emph{minus} the radial distance in the asymptotic  coordinates \eqref{Xsnullradframe}. With these conventions we have\footnote{Eq. \eqref{freeprop} can be shown by a writing the free-field as a  Fourier integral, see e.g. \cite{stromST}. Notice there is no antipodal map on the celestial sphere since we use the same null vector \eqref{defnmu} for both future and past null rays \eqref{Xsnullradframe}. \label{hinconvfnote}}
\be \label{freeprop}
h^\out_{\mu \nu}(u,\nh) = h^\inn_{\mu \nu}(u,\nh) \quad \text{for $O(G^0)$ (free-field) propagation}.
\ee

It is natural to assume, and we will do so, that the Fourier transform of \eqref{defhin}  admits a soft expansion as \eqref{hmunuomega}
\be \label{hmunuinomega}
\tilde{h}^\inn_{\mu\nu}(\omega, \hat{n}) \stackrel{\w \to 0}{=}   \omega^{-1} \htilde^{(0) \inn}_{\mu\nu}(\hat{n}) + \log \omega \, \htilde^{(\log) \inn}_{\mu\nu}(\hat{n})  + \cdots 
\ee

However, in our previous discussion (and generally in the literature)  incoming gravitational radiation, if present at all, is considered to be  purely hard: 
\be \label{noincomingsoft}
 \text{Implicit assumption so far:} \quad  \htilde^{(0) \inn}_{\mu\nu} =  \htilde^{(\log) \inn}_{\mu\nu} =0.
\ee

In presence of non-trivial incoming soft terms, the formulas \eqref{hminuone} and \eqref{hlogdevdrag} acquire additional contributions,
\ba
\htilde^{(0)}_{\mu\nu} &\to&   \htilde^{(0)}_{\mu\nu} + \htilde^{(0) \inn}_{\mu\nu} , \label{hleadincomingsoft} \\
\htilde^{(\log)}_{\mu\nu} &\to&   \htilde^{(\log)}_{\mu\nu} + \htilde^{(\log) \inn}_{\mu\nu}  - 4 i G \, n \cdot P \, \htilde^{(0) \inn}_{\mu\nu}. \label{hlnincomingsoft}
\ea

From a perturbative perspective, the first corrections in \eqref{hleadincomingsoft} and \eqref{hlnincomingsoft} come from the free propagation of the incoming radiation \eqref{freeprop}. The second correction in \eqref{hlnincomingsoft} represents a drag term on the incoming soft radiation, and can be obtained by ``completing" with $\htilde^{(0) \inn}_{\mu\nu}$ the incoming leading soft factor in the second line of  \eqref{hlnasymfield}.\footnote{A naive  guess for this contribution is obtained by substituting \eqref{hleadincomingsoft} in \eqref{hdrag}. The result of this  substitution, however, depends on the log frame in which is performed. In the harmonic log frame where $c^\mu_{\scri^+}=2 G P^\mu$, this  gives half of what is claimed in \eqref{hlnincomingsoft}. The mismatch can be understood from the fact that  \eqref{hdrag} only accounts for outgoing drag, while incoming soft radiation also experiences incoming drag. The naive substitution yields the correct result if performed in the past radiative frame, for which  there is no incoming drag and $c^\mu_{\scri^+}=4 G P^\mu$.} 
All the  terms in \eqref{hleadincomingsoft} and \eqref{hlnincomingsoft} naturally arise in the asymptotic proofs of section  \ref{asymproofsec}.

A benefit of including incoming soft radiation is that it makes manifest the time-reversal symmetry of the soft theorems. Whereas this is straightforward for the leading soft theorem \cite{stromST}, it is less so for the logarithmic one, due to the second line in \eqref{hlnasymfield}. However, upon including the corrections \eqref{hlnincomingsoft} \emph{and} taking into account the leading soft theorem, one can establish the time-reversal covariance of the log soft theorem, see appendix \ref{Tcovapp} for  details.

\section{Spacetime asymptotics} \label{stimeasymssec}

\subsection{Timelike infinity} \label{timesec} 
To describe events near future/past timelike infinity, we consider asymptotic  coordinates $(\t,x^a)$ obtained by setting $s=\t$ and $V^\mu =V^\mu(x)$ in  \eqref{Xstimelike},
\be \label{Xmuitotauxa}
X^\mu(\t,x) \stackrel{\tau \to \pm \infty}{=} \tau V^\mu(x) + \cdots, \quad V^\mu V_\mu = -1, \quad V^0>0,
\ee
with  $V^\mu(x)$ given by Eq. \eqref{Vxtime}. The dots in \eqref{Xmuitotauxa} depend on the  choice of gauge. We adopt Beig-Schmidt (BS) conditions  \cite{BS,cgw},
\be \label{BSgaugepm}
g_{\t a}=0 , \quad g_{\t \t}= -\left(1+\frac{\sigma}{\tau}\right)^2,
\ee
as they allow to separate Einstein equations into a series of elliptic equations on $\H^\pm$ that can be solved recursively \cite{BS,cgw}.

The remaining metric components admit a large $\t$ expansion \cite{cgw}
\be \label{gpmab}
g_{ab} = \tau^2 h_{ab}+ \tau  (k_{ab}-2\sigma h_{ab})+ \log | \tau | \, i_{ab}+O(\t^0),  
\ee
where $h_{ab} = \partial_a V^\mu \partial_b V_\mu$ is the unit hyperboloid metric, used to raise and lower indices $a,b, \ldots$.   In the coordinates $x^a=(\rho,\phi^A)$ \eqref{Vxtime} this metric reads
\be \label{ds2timeH}
h_{ab} dx^a dx^b =\frac{d \rho^2}{\rho^2+1} + \rho^2 d \Omega^2,
\ee
with  $d \Omega^2= \partial_A \nh \cdot \partial_B \nh \, d \phi^A d \phi^B$ the unit sphere line element. The rest of the  coefficients in \eqref{BSgaugepm} and \eqref{gpmab} are   fields on $\H^\pm$ that we  describe below. To avoid clutter, we omit future/past labels on them unless required.\footnote{Our conventions are such that the asymptotic expansions at $\t \to \pm \infty$ look identical. The distinction between future and past coefficients appear when expressing them in terms of asymptotic particles, see e.g. Eq. \eqref{rhomassive}. We note that at $\H^-$ some coefficients carry opposite signs relative to the conventions of \cite{cgw,gianni1}.} 

The scalar $\sigma$ in \eqref{BSgaugepm} plays the role of a  potential for the asymptotic electric part of the Weyl tensor \cite{hansen,dehouck}.  It also captures the log deviation vector according to \cite{gianni1}
\be \label{cpmitosigma}
c^\mu =  D^a V^\mu \partial_a \sigma - V^\mu \sigma.
\ee

The tensor $k_{ab}$ is a potential for the magnetic part of the Weyl tensor \cite{hansen,dehouck} which  vanishes under standard  asymptotic flatness conditions. This implies  it takes a  ``pure gauge'' form 
\be \label{kabitoPhitime}
k_{ab}=-2(D_aD_b-h_{ab})\Phi,
\ee
where $\Phi$ is a scalar that plays the role of a  supertranslations Goldstone mode \cite{phigoldstone}.

The fields of interest for our purposes will be $\sigma$ and $i_{ab}$. $\Phi$ and  $k_{ab}$ play no role in the asymptotic derivation of  soft theorems and thus will  not be discussed any further.

\subsubsection*{Matter stress tensor and Einstein equations}

As explained in section \ref{asympartsec}, the matter content at timelike infinity is described by  massive particles. The resulting stress tensor admits the following large $\t$  expansion (see  appendix \ref{BStimeapp}) 
\be \label{stresstensortime}
\begin{aligned}
T_{\t \t} &=   \frac{1}{\t^3} \rho - \frac{\log |\t|}{\t^4} D \cdot j + O(\t^{-4}) \\
T_{\tau a} &=  \frac{\log |\tau|}{\tau^3} j_a +O(\t^{-3})  \\
 T_{a b} &=  O(\t^{-3} \log^2 |\t|) 
\end{aligned}
 \ee
where  $\rho$  is the (signed) energy density of massive particles at future/past timelike infinity,
\be \label{rhomassive}
\rho\big|_{\H^\pm}(x)=\pm \sum_{\substack{i \in \pm \\ m_i \neq 0}} m_i \delta(x,x_i)
\ee
and
\be \label{defja}
j_a = \rho \, c_a , \quad  c_a = D_a V^\mu c_\mu =  \partial_a \sigma.
\ee

One can show that the expansion \eqref{stresstensortime} is consistent with that of the Einstein tensor, and that Einstein equations  lead to  (see  appendix \ref{BStimeapp}) 
\be \label{eqsigma}
(D^2-3) \sigma = 4 \pi G \rho,
\ee
and
\be \label{iabecstime}
\begin{aligned}
  h^{ab} i_{ab} &=  4\pi G D \cdot j\\
  D^b  i_{\langle ab \rangle} &= 8 \pi G \left(   \tfrac{1}{3}D_a D \cdot j- j_a \right)\\
(D^2+2)i_{\langle ab \rangle} &= 4 \pi G \left(  D_{\langle a} D_{b \rangle} D \cdot j    - 4 D_{\langle a}j_{b\rangle}\right)
\end{aligned}
\ee
where  $i_{\langle ab \rangle}$ is the trace-free part of $i_{ab}$.

In appendix \ref{timegreenapp} we construct the solutions to these elliptic equations by Green's functions methods. In the following we summarize the properties of such solutions that will be  used in the asymptotic proof of the soft theorems.  Before doing so, however, we need to  make a few remarks on log translations.

\subsubsection*{Log translations}
As originally discussed in \cite{BS} in the spatial infinity context, log translations are allowed by conditions \eqref{BSgaugepm}. Their asymptotic form in BS coordinates is
\be
\xi_L  \stackrel{\tau \to \pm \infty}{=} l \ln |\t|  \partial_\t + \cdots, \quad  l := - L^\mu V_\mu,
\ee
where the dots are determined by requiring consistency with  \eqref{BSgaugepm}. By evaluating the Lie derivative on the asymptotic metric and stress tensor one finds\footnote{Eq. \eqref{delLiabtime} only holds when $k_{ab}$ is ``pure gauge'' as in \eqref{kabitoPhitime}, otherwise there is an additional contribution proportional to the magnetic component of the asymptotic Weyl tensor. See Eq. (4.108) of \cite{dehouck}  for the full expression in the spatial infinity case.}
\ba
\delta_L \sigma &= &  l ,\\
 \delta_{L} i_{ab} &=& D^c (D_c l \, (D_a D_b -h_{ab})\sigma),  \label{delLiabtime} \\ 
\delta_{L} j_a &=&  D_a l  \, \rho \label{delLja}, \\
\delta_L h_{ab}&=&  \delta_L k_{ab}= \delta_L \rho =0.
\ea

Note  these transformations are compatible with \eqref{delLcmu},  \eqref{defja} and \eqref{iabecstime}.  For the purposes of matching with null infinity,  log translations will be fixed by imposing future/past radiative log frame conditions.

\subsubsection*{Asymptotic properties of $\sigma$}

Let  $\G(x,x') $ be the Green's function of the differential operator on the LHS of \eqref{eqsigma}, so that
\be \label{sigmaitogreentime}
\sigma(x) = 4 \pi G \int d^3 x' \G(x,x') \rho(x').
\ee
There are potential ambiguities in the definition of  $\G$,  due to homogeneous solutions to  \eqref{eqsigma}, which are  eliminated by the radiative log frame condition. In terms  of $\sigma$, this condition reads \cite{cgw,gianni1}
\be\label{asymsigmacgw}
\lim_{\rho \to \infty}\sigma = 0  \quad \text{(radiative log frame condition}).
\ee
The resulting Green's function is reviewed in appendix \ref{timegreenapp}.   For matching purposes, we will only  need its asymptotic form  at large $\rho$,
\be \label{asymgreensigmatime}
\G(x,x')  \stackrel{\rho \to \infty}{=}  \frac{1}{16 \pi \rho^3}  \frac{1}{(n \cdot V')^3} + \cdots.
\ee
Using \eqref{asymgreensigmatime} and \eqref{rhomassive} in \eqref{sigmaitogreentime} leads to 
\be \label{asyimsigmatime}
\sigma\big|_{\H^\pm}(\rho,\phi)  \stackrel{\rho \to \infty}{=} \frac{\ov{0}{\sigma}\big|_{\partial \H^\pm}(\phi)}{\rho^3}  + \cdots,
\ee
with
\be \label{sigmazerotime}
 \ov{0}{\sigma}\big|_{\partial \H^\pm}=\frac{G}{4}\sum_{\substack{i \in \pm \\ m_i \neq 0}}  \frac{m_i^4}{(p_i \cdot n)^3},
\ee
where  we expressed the result in terms of the asymptotic momenta \eqref{defpinout}.  Note that the $\pm$ signs in \eqref{rhomassive} and \eqref{defpinout} cancel each other so that Eq. \eqref{asyimsigmatime} takes the same form at both infinities.

\subsubsection*{Asymptotic properties of $i_{ab}$}
In appendix \ref{timegreenapp} we show how the last two equations in \eqref{iabecstime} can be solved in terms of a  Green's function $ \G_{ab}^{c'}(x,x')$, 
\be \label{iabitoj}
i_{\langle ab \rangle}(x) = 8 \pi G \int d^3 x' \G_{ab}^{c'}(x,x') j_{c'}(x').
\ee

This time the  log translation ambiguity does not manifest in the Green's function, but in the source term through  Eq. \eqref{delLja}. As before, we fix  this freedom by requiring the radiative log frame condition \eqref{asymsigmacgw}.  

For the matching with null infinity, all we shall need is the asymptotic form of the  radial-sphere components of the Green's function, given by (see appendix \ref{timegreenapp})
\be \label{asymGrhoAbp}
\G_{\rho A}^{b'}(x,x')  \stackrel{\rho \to \infty}{=}  \frac{3}{8 \pi \rho^3}  \frac{ \partial_A n^{\mu} n^{\nu} D^{b'} V'_{[\mu} V'_{\nu]}}{(n \cdot V')^4} + \cdots
\ee

Using \eqref{asymGrhoAbp} and \eqref{defja}   leads to
\be \label{irhoAtime}
i_{\rho A}\big|_{\H^\pm}(\rho,\phi)  \stackrel{\rho \to \infty}{=}  \frac{\izero_{\rho A}\big|_{\partial \H^\pm}(\phi)}{\rho^3} + \cdots,
\ee
with
\be \label{izerotimeinf}
\izero_{\rho A}\big|_{\partial \H^\pm}  = 3 G  \sum_{\substack{i \in \pm \\ m_i \neq 0}} m_i^4 \frac{  \partial_A n^{\mu} n^{\nu} c^i_{[\mu} \, p^i_{\nu]}}{(p_i \cdot n)^4} .
\ee \\

\noindent \emph{Comment} \\
 It is interesting to note the following parallel between the expressions for $\sigma$ and $i_{ab}$: The asymptotic values of the Green's functions \eqref{asymgreensigmatime} and   \eqref{asymGrhoAbp} can be interpreted as boundary-to-bulk Green's function for supertranslations and superrotations respectively \cite{mcgreen}. At the same time,  the asymptotic coefficients \eqref{sigmazerotime}  and \eqref{izerotimeinf} can be interpreted as the particles' contribution to the  mass and angular momentum aspects respectively (the latter for the case of an orbital angular momentum $J_{\mu \nu}=c_{[\mu} \, p_{\nu]}$, see e.g. \cite{Compere:2019gft}). This double interpretation underlies the symmetry realization of the leading and subleading soft graviton theorems in presence of massive particles \cite{clmassive,chipum2}.

\subsection{Spatial  infinity } \label{spisec}
To describe events near spatial infinity, we consider asymptotic  coordinates $(\rho,x^a)$ such that
\be \label{Xmuitorhoxa}
X^\mu(\rho,x)  \stackrel{\rho \to \infty}{=}   \rho \, V^{\mu}(x)  +\cdots,  \quad  V^\mu V_\mu = 1 ,
\ee
where $\rho$ is an asymptotic proper distance and $x^a=(\t,\phi^A)$ parametrize unit spacelike directions according to  \eqref{Vxspatial}.  To extend these coordinates beyond leading order we again adopt Beig-Schmidt conditions \cite{BS},
\be \label{BSgaugespatial}
g_{\rho a}=0 , \quad g_{\rho \rho}= \left(1+\frac{\sigma}{\rho}\right)^2,
\ee
\be \label{gspatialab}
g_{ab}   = \rho^2 h_{ab}+ \rho (k_{ab}-2\sigma h_{ab})+ \log \rho \, i_{ab}+O(\rho^0) ,
\ee
where the  coefficients  are now tensor fields on $\H^0$, with  metric $h_{ab} = \partial_a V^\mu \partial_b V_\mu$ given by 
\be \label{ds2spiH}
h_{ab} dx^a dx^b = -\frac{d\tau^2}{1+\tau^2}+(1+\tau^2) d \Omega^2.
\ee

As before, $\sigma$ and $k_{ab}$ are potentials for the  electric and magnetic parts of the asymptotic Weyl tensor, respectively. The former captures the log deviation vector according to 
\be \label{cpmitosigmaspi}
c^\mu = -D^a V^\mu \partial_a \sigma - V^\mu \sigma,
\ee
while the latter takes the form
\be \label{kabitoPhispi}
k_{ab}=-2(D_aD_b+h_{ab})\Phi,
\ee
due to the vanishing of the asymptotic magnetic Weyl tensor.

We note that one can formally map  the Beig-Schmidt expressions at  timelike infinity to those at spatial infinity  
by doing the replacements \cite{cgw}\footnote{Along with  $ \rho \to - i  \sqrt{\t^2+1}$,  $ V^\mu \to - i  V^\mu$.} 
\be \label{analiticcont}
\t \to i \rho, \quad \sigma \to  i \sigma, \quad h_{ab} \to -h_{ab}, \quad \Phi \to -i \Phi, \quad i_{ab} \to i_{ab}.
\ee

\subsubsection*{Einstein equations}

Einstein equations for $\sigma$ and $i_{ab}$ can  be obtained from those at timelike infinity by applying \eqref{analiticcont}, and setting to zero the source terms.
This leads to
\be \label{eomsigmaspi}
(D^2 +3 ) \sigma =0,
\ee
and
\be \label{eomiabspi}
  h^{ab}i_{ab} = 0 \quad D^b  i_{ ab } = 0 , \quad (D^2-2)i_{ab} = 0.
\ee

In appendix \ref{greenspiapp} we construct the solutions to these hyperbolic equations in terms of initial (final) data at the asymptotic  past (future)  boundary of $\H^0$,  assuming decaying boundary conditions as required for consistency with the Bondi expansion at null infinity \cite{cgw}. After a few comments on log translations, we summarize below the key aspects of such solutions.


\subsubsection*{Log translations}
At spatial infinity, log translations take the asymptotic form
\be
\xi_L   \stackrel{\rho \to \infty}{=}  l \ln \rho  \, \partial_\rho + \cdots, \quad  l :=  L^\mu V_\mu,
\ee
and act on the metric components by \cite{dehouck}
\ba
\delta_L \sigma &= &  l , \label{delLsigmaspi}  \\
 \delta_{L} i_{ab} &=&   D^c \big( D_c l  \, \E_{ab}  \big), \label{deliabspi}  \\ 
\delta_L h_{ab}&=&  \delta_L k_{ab}= 0,
\ea
where
\be \label{Eweyl} 
\E_{ab} = ( D_a D_b + h_{ab})\sigma 
\ee
is the asymptotic electric Weyl curvature \cite{hansen}, itself invariant under log translations.

Of special importance for our analysis is the  log translation that interpolates between future and past radiative frames \eqref{cradpluscradminus}. In this case $L^\mu = 4 G P^\mu  $ and relations \eqref{delLsigmaspi}, \eqref{deliabspi} become\footnote{Although  \eqref{delLsigmaspi} and \eqref{deliabspi} refer to  infinitesimal variations, they also hold for \emph{finite} variations. For \eqref{delLsigmaspi} this property is evident, while for \eqref{deliabspi} it follows because   $\delta_L \E_{ab}=0$.}
\ba
 \sigma^{\rad +} -  \sigma^{\rad -} &= & 4 G P^\mu  V_\mu , \label{difradssigmaspi} \\
 i_{ab}^{\rad +} -  i_{ab}^{\rad -} &=&  4 G P^\mu  D^c \big( D_c V_\mu  \, \E_{ab}  \big), \label{Deltaiab}
\ea
where  $\rad +$/$\rad -$ refers to the solution in the future/past log radiative frame.

\subsubsection*{Asymptotic properties of $\sigma$}

As emphasized in \cite{cgw} it is crucial to distinguish between solutions to \eqref{eomsigmaspi}  in different log frames. In particular, decaying conditions can only be imposed at either the  future or past,
\be \label{asymsigmaspi}
\sigma^{\rad \pm}(\t,\phi) \stackrel{\t \to \pm \infty}{=} \frac{\sigmazero\big|_{\partial_\pm \H^0}(\phi)}{\t^3} + \cdots,
\ee
where the coefficients $\sigmazero\big|_{\partial_\pm \H^0}$  provide final/initial data that parametrize solutions to \eqref{eomsigmaspi} in future/past radiative log frame. For instance, in the future radiative frame (see appendix  \ref{greenspiapp} for details)
\be \label{sigmaradpluslargetau}
\sigma^{\rad +}(x) = \frac{4}{\pi} \int d^2 \phi' \, V(x) \cdot n(\phi') \theta \big(V(x) \cdot n(\phi')\big)  \sigmazero\big|_{\partial_+ \H^0}(\phi'),
\ee
where $\theta$ is the step function.  Analogous expression holds for $\sigma^{\rad -}$ in terms of $\sigmazero\big|_{\partial_- \H^0}$. The two are however not independent. 
 Using the relation\footnote{This expression can be obtained by evaluating the general formula  \cite{hansen,romano,mmvirmani} $P^\mu= (1/8 \pi G) \oint \E^{ab} \partial_b V^\mu d S_a$ on a $\t \to \infty$ slice of $\H^0$, see also \cite{tn,magnonenergymom}.} 
\be \label{Pmuitosigmazero}
P^\mu = \frac{1}{\pi G} \int d^2 \phi \, n^\mu(\phi)  \sigmazero\big|_{\partial_+ \H^0}(\phi),
\ee
one can verify that the decaying solution at $\t \to -\infty$ can be obtained from \eqref{difradssigmaspi} and satisfies
\be \label{matchingsigmazero}
 \sigmazero\big|_{\partial_- \H^0} =- \A_*  \sigmazero\big|_{\partial_+ \H^0},
\ee
where $\A$ is the antipodal map \eqref{defantipodalmap}. This well-known result  \cite{cedric,prabhu,capone,cgw}   underlies the matching of the Bondi mass aspect at spatial infinity \cite{herber,stromgravscatt} that we shall encounter in Eq. \eqref{matchingmassspi}.\\

\noindent \emph{Comment} \\
Combining \eqref{cpmitosigmaspi} and \eqref{sigmaradpluslargetau} one obtains an integral expression for the  log deviation (in future radiative frame) 
\be \label{cplusitosigmazero}
c^{\rad +}_\mu(x) = - \frac{4}{\pi} \int d^2 \phi' \, n_\mu(\phi') \, \theta \big(V(x) \cdot n(\phi')\big)   \sigmazero\big|_{\partial_+ \H^0}(\phi').
\ee
Using \eqref{Pmuitosigmazero} one can verify \eqref{cplusitosigmazero} satisfies \eqref{obstructioncmuspi}. Similar considerations apply  for the log deviation vector in other frames. 

\subsubsection*{Asymptotic properties of $i_{ab}$}

We now consider solutions to \eqref{eomiabspi} in terms of initial/final data,
\be \label{itauAspi}
i_{\t A}(\t,\phi) \stackrel{\t \to \pm \infty}{=}  \frac{\izero_{\t A}\big|_{\partial_\pm \H^0}(\phi)}{\t^3} + \cdots.
\ee

Unlike the previous case, the Green's function for this problem is insensitive to log translation ambiguities. These instead manifest in  the boundary values $\izero_{\t A}\big|_{\partial_\pm \H^0}$, see below. In appendix \ref{greenspiapp} we discuss the general solution of \eqref{eomiabspi}  under \eqref{itauAspi} and show it satisfies 
\be \label{antipodalmatchingitauA}
\izero_{\t A}\big|_{\partial_- \H^0}= -\A_* \izero_{\t A}\big|_{\partial_+ \H^0}.
\ee

For the asymptotic proof of the soft theorem,  we will need to compare the future/past asymptotic values of the $\rad +$/$\rad -$ log frames expressions for $i_{\t A}$. To this end,   consider the large $\t$ expansion of \eqref{deliabspi}, which we write as
\be
\delta_L i_{\t A}(\t,\phi) \stackrel{\t \to \pm \infty}{=}  \frac{\delta_L \izero_{\t A}\big|_{\partial_\pm \H^0}(\phi)}{\t^3} + \cdots,
\ee
with (see the end of appendix \ref{greenspiapp})\footnote{We note that  \eqref{delLitauAplus} is compatible with \eqref{antipodalmatchingitauA} thanks  \eqref{matchingsigmazero}.}
\bal \label{delLitauAplus}
\delta_L\izero_{\t A}\big|_{\partial_+ \H^0} &=  -4(  n \cdot L \partial_A + 3  \partial_A n \cdot L ) \sigmazero\big|_{\partial_+ \H^0} ,\\
\delta_L \izero_{\t A}\big|_{\partial_- \H^0} &=  -4(  \A_* n \cdot L \partial_A + 3  \partial_A \A_* n \cdot L ) \sigmazero\big|_{\partial_- \H^0} .
\eal
  The large $\t$ limit of Eq.  \eqref{Deltaiab} then implies
\be \label{matchingitauAspi}
\izero^{\rad +}_{\t A}\big|_{\partial_+ \H^0}     =-  \A_* \izero^{\rad -}_{\t A}\big|_{\partial_- \H^0} - 16 G  (  n \cdot P \partial_A + 3  \partial_A n \cdot P ) \sigmazero\big|_{\partial_+ \H^0} .
\ee

This ``matching'' 
property will allow us to relate the logarithmic angular momentum aspects at future and past infinity in Eq. \eqref{Ncalspi}. In contrast to Eq. \eqref{matchingsigmazero} for  $\sigma$, there is now a inhomogeneous term that captures a discontinuity across spatial infinity. In the analysis of section \ref{asymproofsec}, this term will generate  the ``extra'' contribution  to the soft factor given in the second line of \eqref{hlnasymfield}.

\subsection{Null infinity} \label{nullsec}

To describe events near future/past null infinity, we consider asymptotic  coordinates $(r,u,\phi^A)$, obtained by setting $s=r$ in \eqref{Xsnullradframe}
\be  \label{Xmunullsec}
X^\mu(r,u,\phi)  \stackrel{r \to \pm \infty}{=} r \, n^\mu(\phi)  + u \, t^\mu + \cdots .
\ee
To fix the dots in \eqref{Xmunullsec} we adopt Bondi-Sachs gauge conditions  \cite{bondi,sachs} 
\be \label{bondisachsgge}
g_{rr}=g_{rA}=0, \quad \det g_{AB} = r^4 \det q_{AB},
\ee
where $q_{AB}$ is the unit sphere metric, used to raised and lower $2d$ indices.  Note that our definition of radial coordinate $r$ is such that it is positive/negative for future/past coordinates; depending on the case,  $u$ represents either retarded or advanced time.  These conventions will allow us to have identical expressions at future and past null infinities. 

 We assume Winicour's logarithmic  asymptotic flatness conditions \cite{winicour} for the large $r$ expansion of the  metric components, as required for generic scattering spacetimes \cite{damour,geillerpeeling,compererobert}:\footnote{The  fall-offs of subleading terms are taken from \cite{geillerpeeling} but their precise form is not important in our analysis.}
\bal
g_{uu} & = -1 + \frac{2 G  \M}{r} + O(\ln r /r^2) \\
g_{ur} &= -1 + O(1/r^2) \\
g_{AB} &= r^2 q_{AB} +r \, C_{AB} + O(r^0) \\
g_{uA} &=  \frac{1}{2} D^B C_{AB}+  \log |r| \frac{2 G }{3 r} \, \ov{\log r}{\N_A}+  \frac{2 G }{3 r } \Nt_A + O(\ln^2 r/r^2). \label{bondimetric}
\eal
The coefficients in this expansion are  regarded as fields on $\I^\pm$, i.e. as functions of $(u,\phi^A)$, and have the following properties:

\begin{itemize}

\item Bondi's shear  $C_{AB}$ is trace-free and  unconstrained by Einstein equations. Its time derivative is denoted by
\be \label{defnews}
N_{AB}=  \partial_u C_{AB} ,
\ee
and referred to as the news tensor.
\item The mass aspect  $\M$ is constrained by Einstein equations to satisfy
\be \label{dotmassaspecteq}
\partial_u \M =    \frac{1}{4 G} D_A D_B N^{AB}  - 4 \pi \rho_{\massless},
\ee
where  $\rho_{\massless}$ is the  energy flux at future/past null infinity, including that due to  gravitational radiation,
\be
 \rho_{\grav } = \frac{1}{32 \pi G} N_{AB} N^{ AB}.
\ee

\item The coefficient $\Nt_A $ contains in it the angular momentum aspect $\N_A$ (i.e. the angular momentum angular density at  given $u$). There are various prescriptions in the literature for such  quantity \cite{compnich}. For our purposes we find it convenient to define it according to
\be
 \Nt_A =  \N_A +u \partial_A \M -\frac{3}{32 G}\partial_A C^2 +\frac{u}{4 G} D^{B} D_{[B}D^{C}C_{A]C},
\ee
where $C^2\equiv C^{AB} C_{AB}$.   This definition coincides with that of \cite{HPS,cgw}, except for the last $O(u\, C/G)$ term. With this piece added, Einstein equations (written  for simplicity  in holomorphic  coordinates \eqref{nhitozzb}) imply 
\be \label{dotangmomasp}
\partial_u {\N_z} = - \frac{1}{2G} u   D_z^3 {N^{ zz}} + \text{angular momentum flux},
\ee
where the ``angular momentum flux'' can be due to  massless matter or gravitational radiation, the latter being quadratic in the shear/news.\footnote{Consisting in terms of the form  $ O(C N/G)$ and $ O(u N^2/G)$.  The angular momentum flux  is needed for the asymptotic description of the tree-level \(O(\w^0)\) soft theorem, as it  accounts for the massless ``hard'' contribution \cite{stromvirasoro}. It is, however, irrelevant for the \(O(\log \w)\) soft theorem and this is why we do not display it explicitly; see  Eq. (2.6) of \cite{HPS} for its expression.}

\item  The coefficient $\ov{\log r}{\N_A}$ is an extra component allowed by Winicour that is excluded  in the original work of Bondi and Sachs. It may be thought of as a   ``divergent in $r$'' angular momentum aspect.  Consistency with Einstein equations requires it is  $u$-independent
\be
\partial_u \ov{\log r}{\N_A} =0.
\ee
This field comes along with a $O(r^{0})$ trace-free component in $g_{AB}$ \cite{sachs,winicour,BT} that will not be needed  in the present work.  

\end{itemize}

\subsubsection*{Large-$u$ behavior}

To fully characterize the spacetime metric at null infinity, we need to specify the large $u$  behavior of the shear. Compatibility with the soft expansion \eqref{hmunuomega}  requires \cite{laddhasen}\footnote{We omit for now   the labels that distinguish the coefficients at  $u \to \pm \infty$.}
\be \label{largeushear}
C_{AB}(u,\phi) \stackrel{u \to  \pm \infty}{=}  \ov{0}{C}_{AB}(\phi)  + \frac{1}{u} \ov{1}{C}_{AB}(\phi)  + \cdots,
\ee
with a ``purely electric'' leading term \cite{stromgravscatt,stromST}
\be \label{electricCzero}
D_{[A}D^{C}\ov{0}{C}_{B]C} =0.
\ee

Eq. \eqref{largeushear} implies the  news tensor  falls off as $1/u^2$ and 
\be \label{falldotMdotNz}
\partial_u \M  =  O(1/u^2) , \quad  \partial_u \N_z  =  O(1/u) ,
\ee
where the decaying rates in \eqref{falldotMdotNz} are dictated by the linear-in-news terms in \eqref{dotmassaspecteq} and \eqref{dotangmomasp} respectively.  Integrating  \eqref{falldotMdotNz} in $u$ leads to
\ba
\M(u,\phi)& \stackrel{ u \to \pm   \infty}{=} &\ov{0}{\M}(\phi) + O(1/u) \label{largeumassasp} \\
\N_z(u,\phi)& \stackrel{u \to \pm   \infty}{=}&  \log | u | \, \ov{\log u}{\N_z}(\phi)    + O(u^0),  \label{largeuangasp}
\ea
where\footnote{We note that, because of \eqref{electricCzero}, Eq. \eqref{NlnitoCone} holds unchanged for the angular momentum aspect of \cite{cgw}.}
\be \label{NlnitoCone}
\ov{\log u}{\N_z}=  \frac{1}{2 G} D_z^3 \ov{1}{C}^{zz},
\ee
and where the difference of the $\ov{0}{\M}$ coefficients at $u \to \pm \infty$  is restricted by the total $u$-integral of \eqref{dotmassaspecteq}.

\subsubsection*{Logarithmic angular momentum aspect}
The Bondi metric \eqref{bondimetric} exhibits two distinct logarithmic divergences in the angular momentum aspect, captured by $\ov{\log r}{\N_A}$ and  $\ov{\log u}{\N_A}$.
For the discussion of the upcoming sections, it will be useful to combine these two terms in a single ``log angular momentum aspect'' defined by
\be \label{logangmomasp}
\ov{\log }{\N_A}:=\ov{\log r}{\N_A}+  \ov{\log u}{\N_A}.
\ee

The proof of the soft theorem is actually insensitive to the value of $\ov{\log r}{\N_A}$, and one may therefore be tempted to take it to  zero. However, consistency with the individual $u \to \pm \infty$ coefficients of  $\ov{\log u}{\N_A}$ requires a non-trivial $\ov{\log r}{\N_A}$ \cite{gianni3}. 

 We finally note  that both logarithmic divergences  vanish upon 2d   smearing with Lorentz generators and hence they are  consistent with a  finite total angular momentum. For the $\log u$ term, this follows from \eqref{NlnitoCone} while for the $\log r$ term this is because it can be written as the divergence of a symmetric, traceless 2d tensor \cite{winicour}.

\subsection{Matching conditions} \label{matchingsec}

\begin{figure}[ht]
  \centering
  \includegraphics[width=0.4\textwidth]{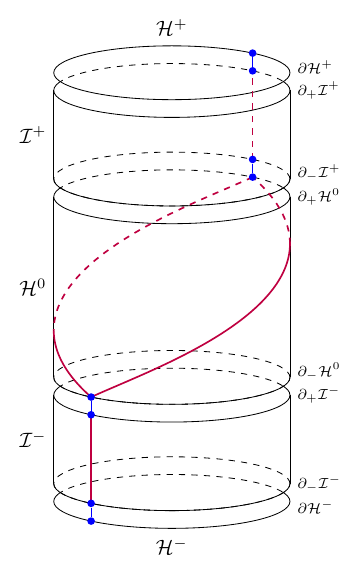}
  \caption{Simplified drawing of the five infinities and their boundaries, together with  a null ray at infinity that connects asymptotic past and future null directions. The hyperbolic spaces $\H^\pm$ are depicted as disks, and the de Sitter space $\H^0$ as a  cylinder.   The blue dots represent  points at each of the eight boundaries.   There are four blue links indicating their matching across adjacent boundaries.  
  The red vertical lines are null generators at $\I^\pm$ . The curved red lines  represent the evolution of  null rays across $\H^0$.}
  \label{figure2}
\end{figure}

In the previous subsections we described the spacetime metric near each of the  five  asymptotic boundaries depicted in the right panel of Figure \ref{figure1},
\be \label{5bdies}
\H^- , \scri^- , \H^0, \scri^+ ,\H^+ .
\ee
In order to complete the  CGW framework, we need to describe the matching conditions at the adjacent boundaries of  \eqref{5bdies}  (see also Figure \ref{figure2})
\bal \label{bdiesident}
\partial \H^- \sim \partial_- \I^- , \quad  \partial_+ \I^-  \sim \partial_- \H^0, \quad \partial_+ \H^0 \sim \partial_- \I^+, \quad  \partial_+ \I^+ \sim \partial \H^+.
\eal

Below we  discuss the matching properties that are needed for the present work. A sketch of their proof is given in appendix \ref{matchingapp}, and we refer the reader to \cite{cgw} for further details.

Consistency between the Beig-Schmidt and Bondi expansions leads to the identification of the boundary values of the BS potential  (Eqs. \eqref{asyimsigmatime} and \eqref{asymsigmaspi})  and Bondi mass aspect  \eqref{largeumassasp} according to
\ba \label{sigmamasstime}
\sigmazero^{\rad \pm}\big|_{\partial \H^\pm} &=& - \frac{G}{4}  \Mzero\big|_{\partial_\pm \scri^\pm}, \\
\sigmazero^{\rad +}\big|_{\partial_+ \H^0} &= &    \phantom{-}\frac{G}{4}  \Mzero\big|_{\partial_- \I^+}, \label{sigmamassspip} \\
\sigmazero^{\rad -}\big|_{\partial_- \H^0} &=  &  \phantom{-} \frac{G}{4} \A_* \Mzero\big|_{\partial_+ \I^-}. \label{sigmamassspim}
\ea
where we have now made explicit all labels. 

Similarly, the leading component of the logarithmic BS metric component \eqref{irhoAtime} and \eqref{itauAspi} can be identified with the  logarithmic angular momentum aspect \eqref{logangmomasp} according to
\ba
 \ov{0}{i}^{\rad \pm}_{\rho A}\big|_{\partial \H^\pm}& =&  -G \, \ov{\log }{\N_A}\big|_{\partial_\pm \scri^\pm} , \label{irhoAangasptime} \\
\ov{0}{i}^{\rad +}_{\t A}\big|_{\partial_+ \H^0} &= & \phantom{-}G \,  \ov{\log }{\N_A}\big|_{\partial_- \I^+} , \label{itAitoNAspip}   \\
\ov{0}{i}^{\rad -}_{\t A}\big|_{\partial_- \H^0} &= & \phantom{-} G  \A_*  \ov{\log }{\N_A}\big|_{\partial_+ \I^-} \label{itAitoNAspim} .
\ea

\section{Asymptotic proof of soft theorems} \label{asymproofsec}

We finally show how the results from the previous section imply the leading and log  soft theorems, reviewed in section \ref{softthmsec}. The general idea behind  these  \emph{asymptotic proofs} is illustrated in Figure \ref{figure2}. The direction of the soft radiation is associated to a null ray that leaves its imprint at all  five infinities. Using Einstein equations at each of these 3d hypersurfaces, together with  matching conditions across their 2d boundaries, allows one to evaluate the soft component of the radiation in terms of energy fluxes at timelike and null infinities.  

For the leading soft theorem this type of analysis has already been discussed in \cite{stromST,cedric,prabhu,capone,cgw}. The purpose of its exposition here is to present the overall rationale   in the simplest setting, as well as to highlight the new features that appear in the log case.

The results we are about to discuss can additionally be interpreted as conservation laws for asymptotic charges \cite{stromST,chipum2}. Here we will not rely on this interesting perspective but just limit ourselves to establish the connection between Einstein equations, matching conditions and soft theorems.

\subsection{Soft factors in Bondi variables}

The first step is to express the soft theorems  in terms of the Bondi fields of section \ref{nullsec}.  The outgoing/incoming Cartesian waveform \eqref{defhout}/\eqref{defhin} is related to the future/past Bondi shear by
\be \label{Citoh}
{C_{AB}|_{\scri^\pm}} = \partial_A n^\mu \partial_B n^\nu h_{\mu \nu}|_{\scri^\pm} 
\ee
where $h_{\mu \nu}|_{\scri^+} \equiv h^\out_{\mu \nu} $ and $h_{\mu \nu}|_{\scri^-} \equiv h^\inn_{\mu \nu}$.  The soft-frequency coefficients in \eqref{hmunuomega} and \eqref{hmunuinomega} can then be written in terms of differences of the large $u$ coefficients \cite{senreview} of the Bondi shear \eqref{largeushear},
\ba
 \big[ {\ov{0}{C}_{AB}\big]_{\scri^\pm}} &=&   -  i  \partial_A n^\mu \partial_B n^\nu \htilde^{(0)}_{\mu\nu}|_{\scri^\pm} ,  \label{delCzero} \\
 \big[ {\ov{1}{C}_{AB}\big]_{\scri^\pm}}  &=&    -  \partial_A n^\mu \partial_B n^\nu \htilde^{(\log)}_{\mu\nu}|_{\scri^\pm}   ,\label{delCone}
\ea
where 
\be \label{defDeltaCn}
 \big[ {\ov{n}{C}_{AB}\big]_{\scri^\pm}}:= {\ov{n}{C}_{AB}|_{\partial_+\scri^\pm}}- {\ov{n}{C}_{AB}|_{\partial_-\scri^\pm}}, \quad n=0,1.
\ee 

For the proof of the soft theorems, we will relate these coefficients to the mass and angular aspects respectively.

\subsection{Leading soft theorem} \label{derivationleadingthm} 
In terms of the Bondi coefficients \eqref{defDeltaCn}, the leading soft theorem \eqref{hminuone} (including  incoming memory according to Eq. \eqref{hleadincomingsoft}) takes the form
\be \label{leadingthmbondishear}
 \big[ {\ov{0}{C}_{AB}\big]_{\scri^+}}= - 4 G  \partial_A n^\mu \partial_B n^\nu \sum_i \frac{p^i_\mu p^i_\nu}{p_i \cdot n} + \big[ {\ov{0}{C}_{AB}\big]_{\scri^-}} .
\ee

In order to relate this expression to the mass aspect, we  take the double sphere divergence of \eqref{leadingthmbondishear}.   Using the identities  reviewed in appendix \ref{2didsapp} this leads to
 \be  \label{leadingthmrew}
 D^A D^B \big[ {\ov{0}{C}_{AB}\big]_{\scri^+}}= - 4 G \sum_{i} \frac{m_i^4}{(p_i \cdot n)^3} +  D^A D^B  \big[ {\ov{0}{C}_{AB}\big]_{\scri^-}},
\ee
where  the sum includes both massive and massless particles,   the latter understood by a  $m_i \to 0$ limit  (see below). So far, Eq.  \eqref{leadingthmrew} is just  a rewriting of the leading soft theorem.\footnote{The original  version   \eqref{leadingthmbondishear} can be recovered from \eqref{leadingthmrew}  and \eqref{electricCzero} \cite{stromST}.}  We now show how this relation follows from the asymptotic Einstein equations and  the matching conditions.

We start by noting that Eqs.   \eqref{matchingsigmazero}, \eqref{sigmamassspip}, \eqref{sigmamassspim}   lead to  Strominger's identification of the Bondi mass aspects at spatial infinity \cite{stromgravscatt}, which in our conventions reads
\be \label{matchingmassspi}
\Mzero|_{\partial_-\scri^+} = - \Mzero|_{\partial_+\scri^-} .
\ee

The next step is to express each side of  \eqref{matchingmassspi} in terms of data at null and timelike infinity. To this end, consider the integrated version of \eqref{dotmassaspecteq}
\be \label{deltabondimass}
\Mzero|_{\partial_+\scri^\pm}- \Mzero|_{\partial_-\scri^\pm}=    \frac{1}{4 G} D^A D^B  \big[ {\ov{0}{C}_{AB}\big]_{\scri^\pm}} - 4 \pi \int du \rho_{\massless}\big|_{\scri^\pm}.
\ee
The mass aspect at timelike infinity  can be evaluated by combining Eqs. \eqref{sigmamasstime} and \eqref{sigmazerotime},
\be \label{massaspecttime}
\Mzero|_{\partial_\pm \scri^\pm}=-\sum_{\substack{i \in \pm \\ m_i \neq 0}}  \frac{m_i^4}{(p_i \cdot n)^3}.
\ee
Meanwhile, the  massless radiated  energy per solid angle   can be written as  \cite{stromgravscatt,thorne,zhibo},
\be \label{rhomasslessparticle}
\int d u  \rho_{\massless}(u,\phi)\big|_{\scri^\pm} =  \pm \sum_{\substack{i  \in \pm \\ m_i = 0}} E_i  \delta( \phi,\phi_i)  .
\ee
where $(E_i,\phi_i)$ parametrize the  null momentum   $p^\mu_i = E_i n^\mu(\phi_i)$. 

Solving for $\Mzero|_{\partial_\mp \scri^\pm}$ in  \eqref{deltabondimass} and using Eqs. \eqref{massaspecttime}, \eqref{rhomasslessparticle} leads to
\ba
\Mzero|_{\partial_\mp \scri^\pm} & = & \mp  \frac{1}{4 G} D^A D^B  \big[ {\ov{0}{C}_{AB}\big]_{\scri^\pm}} -\sum_{\substack{i \in \pm \\ m_i \neq 0}}  \frac{m_i^4}{(p_i \cdot n)^3}  + 4 \pi   \sum_{\substack{i \in \pm  \\ m_i = 0}} E_i \delta(\nh,\ph_i) \\
& = & \mp  \frac{1}{4 G} D^A D^B  \big[ {\ov{0}{C}_{AB}\big]_{\scri^\pm}} -\sum_{i \in \pm}  \frac{m_i^4}{(p_i \cdot n)^3} , \label{Mzerospi}
\ea
where in the second line we combined   massive and massless contributions by means of the identity \cite{clmassive}
\be
\lim_{m_i \to 0}  \frac{m_i^4}{(p_i \cdot n(\phi))^3}  = - 4 \pi E_i \delta(\phi,\phi_i).
\ee

Substituting \eqref{Mzerospi} in \eqref{matchingmassspi} and solving for $ D^A D^B  \big[ {\ov{0}{C}_{AB}\big]_{\scri^+}}$ leads to \eqref{leadingthmrew}.

\subsection{Log soft theorem} \label{derivationlogthm}

As before, we start by expressing the log soft theorem, including incoming memory \eqref{hlnincomingsoft}, in terms of the subleading Bondi coefficient \eqref{defDeltaCn},
\be \label{subleadingthmbondishear}
 \big[ {\ov{1}{C}_{zz}\big]_{\scri^+}}= -   \partial_z n^\mu \partial_z n^\nu  \htilde^{(\log)}_{\mu\nu}   + 4 G \, n \cdot P \big[ {\ov{0}{C}_{zz}\big]_{\scri^-}} + \big[ {\ov{1}{C}_{zz}\big]_{\scri^-}},
\ee
where $\htilde^{(\log)}_{\mu\nu}$ stands for the expression \eqref{hlnasymfield} and we choose 2d holomorphic coordinates  to simplify the analysis.

Let us further rewrite \eqref{subleadingthmbondishear}  in terms of differences of the logarithmic  angular momentum aspect \eqref{logangmomasp},
\be \label{difNlogeqDz3C1}
 \big[ \ov{\log }{\N}_{z}\big]_{\scri^\pm}=     \frac{1}{2 G} D_z^3  \big[ {\ov{1}{C}^{zz}\big]_{\scri^\pm}},
\ee
where we used Eq. \eqref{NlnitoCone} together with the fact that  $\big[ \ov{\log r}{\N}_{z}\big]_{\scri^\pm}=0$ due to its $u$-independence.

Using the identities presented in Appendix \ref{2didsapp}, the third derivative in \eqref{difNlogeqDz3C1} can be evaluated on the various terms in \eqref{subleadingthmbondishear} leading to 
\begin{multline}  \label{logsoftthmrewproof}
 \big[ \ov{\log}{\N}_{z}\big]_{\scri^+} = -3  \sum_{\substack{i  \in \out \\ m_i \neq 0}}  m_i^4 \frac{  \partial_z n^{\mu} n^{\nu}    J^{\rad +i}_{\mu \nu}  }{(p_i \cdot n)^4} -  3  \sum_{\substack{i  \in \inn \\ m_i \neq 0}}  m_i^4 \frac{  \partial_z n^{\mu} n^{\nu}    J^{\rad - i}_{\mu \nu}  }{(p_i \cdot n)^4} \\
 +    \left(  n \cdot P  \partial_z  + 3 \partial_z  n \cdot P \right)  \left( 4 G \sum_{i  \in \inn} \frac{m^4_i}{(p_i \cdot n)^3} -   2 D^2_z   \big[ {\ov{0}{C}^{zz}\big]_{\scri^-}}   \right)\\
 +  \big[ \ov{\log }{\N}_{z}\big]_{\scri^-} .
\end{multline}
This expression   provides  a reformulation of the log soft theorem, analogous to  \eqref{leadingthmrew} for the leading soft theorem.\footnote{One can recover \eqref{subleadingthmbondishear} from \eqref{logsoftthmrewproof} by inverting the $D_z^3$ differential operator \cite{stromvirasoro,relaxed}.} 
Our goal now is to show that this identity follows from the matching properties of the asymptotic fields.

Let us start by  considering the  logarithmic angular momentum aspect at spatial infinity.  Combining Eqs.  \eqref{matchingitauAspi}, \eqref{sigmamassspip},   \eqref{itAitoNAspip}, \eqref{itAitoNAspim} and \eqref{matchingmassspi} leads to
\be \label{Ncalspi}
  \ov{\log }{\N}_{z}\big|_{\partial_- \scri^+} =-    \ov{\log }{\N}_{z}\big|_{\partial_+ \scri^-} + 4 G  (  n \cdot P \partial_z + 3  \partial_z n \cdot P )  \Mzero|_{\partial_+ \I^-} .
\ee
This equation is the analogue of \eqref{matchingmassspi} for the Bondi mass aspect, except that now  there is a discontinuity captured in the last term.

Next, we express  \eqref{Ncalspi} in terms of data at null and timelike infinity.  The logarithmic angular momentum aspect at spatial infinity can be written as
\ba
 \ov{\log }{\N}_{z}\big|_{\partial_\mp \scri^\pm} &=& \mp  \big[ \ov{\log}{\N}_{z}\big]_{\scri^\pm} +   \ov{\log }{\N}_{z}\big|_{\partial_\pm \scri^\pm} \\
&=& \mp  \big[ \ov{\log}{\N}_{z}\big]_{\scri^\pm}   - 3 \sum_{\substack{i \in \pm \\ m_i \neq 0}} m_i^4 \frac{  \partial_z n^{\mu} n^{\nu} J^{\rad \pm i}_{\mu \nu} }{(p_i \cdot n)^4}, \label{angaspdata}
\ea
where in the last equality we used  Eqs. \eqref{izerotimeinf} and \eqref{irhoAangasptime} with $ J^{\rad \pm i}_{\mu \nu} \equiv c^{\rad \pm i}_{[\mu} \, p^i_{\nu]}  $.  The  last term in \eqref{Ncalspi} can be evaluated from \eqref{Mzerospi},
\be \label{Mzeroin}
\Mzero|_{\partial_+\scri^-}= -\sum_{i \in \inn}  \frac{m_i^4}{(p_i \cdot n)^3} +    \frac{1}{2 G}   D^2_z \big[ \ov{0}{C}^{zz}\big]_{\scri^-}.
\ee
where we used \eqref{electricCzero} to write  $ D^A D^B  \big[ \ov{0}{C}_{AB}\big]_{\scri^-}= 2 D^2_z   \big[ \ov{0}{C}^{zz}\big]_{\scri^-}$.  Substituting \eqref{angaspdata} and \eqref{Mzeroin}  in \eqref{Ncalspi} and solving for  $\big[\ov{\log}{\N}_{z}\big]_{\scri^+}$ leads to \eqref{logsoftthmrewproof}.\\

 \acknowledgments
We thank Federico Capone, Geoffrey Compère, Rodrigo Eyheralde, Alok Laddha,  Guzmán Hernández-Chifflet and Sébastien Robert for fruitful discussions and Federico Capone for feedback on the draft.  We acknowledge  support from Pedeciba and from ANII grants POS-NAC-2023-1-177577 and FCE-1-2023-1-175902.

\appendix

\section{Time-reversal covariance of soft theorems} \label{Tcovapp}

In a scattering context, time-reversal  symmetry exchanges outgoing and incoming data through the map $t \mapsto -t$ between future and past asymptotic Cartesian times. Denoting this symmetry by $\T$ and with the conventions of Eqs.~\eqref{defhout} and \eqref{defhin}, one has\footnote{In this appendix, indices $i,j,\ldots$ are reserved to denote spatial Cartesian components.} 
\bal \label{Thout}
\T h^\out_{i j}(u, \nh) &= & -  h^\inn_{i j}(-u, -\nh)  \\ 
  \T h^\out_{0 i}(u, \nh)&= &  h^\inn_{0 i }(-u, -\nh) \\
     \T h^\out_{0 0}(u, \nh)&=&  - h^\inn_{00}(-u, -\nh),
\eal
with analogue  expressions  for  $\T h^\inn_{\mu \nu}$. 

Asymptotic momenta change by reversing their spatial components.  Under the conventions of \eqref{defpinout} this leads to
  \be \label{Tp}
\T p^\mu_\out = (-p^0_\inn, \vec{p}_\inn), \quad \quad \T p^\mu_\inn = (-p^0_\out, \vec{p}_\out).
\ee

For definitiveness  we will discuss the purely spatial components of the soft theorems, but similar considerations go through for the  remaining  components.  Eq. \eqref{Thout} implies the spatial components of the soft frequency coefficients transform as
  \be \label{Th0}
\T  \htilde^{(0) \out}_{i j}(\nh)  =  \htilde^{(0) \inn}_{i j}(-\nh) ,  \quad    \T  \htilde^{(0) \inn}_{i j}(\nh) = \htilde^{(0) \out}_{i j}(-\nh)   .
\ee
and
\be \label{Thlog}
\T  \htilde^{(\log) \out}_{i j}(\nh)  =  - \htilde^{(\log) \inn}_{i j}(-\nh) , \quad    \T  \htilde^{(\log) \inn}_{i j}(\nh) = - \htilde^{(\log) \out}_{i j}(-\nh)  .
\ee

As a warmup, consider first the  leading soft theorem,
 \be \label{leadthmTcovapp}
\frac{1}{4Gi}\htilde^{(0) \out}_{i j}(\nh) +   \sum_{\out} \frac{p_i p_j}{p \cdot n} =  \frac{1}{4Gi}\htilde^{(0) \inn}_{i j}(\nh) -  \sum_{\inn}  \frac{p_i p_j}{p \cdot n},
\ee
where we  separated  incoming and outgoing terms and  we left implicit the particle index   to avoid confusion with the spatial indices.   Using \eqref{Tp} and \eqref{Th0}, the time-reversal of \eqref{leadthmTcovapp} is
\be \label{Tleadthm}
\T \eqref{leadthmTcovapp} : \quad  \frac{1}{4Gi}\htilde^{(0) \inn}_{i j}(-\nh) -   \sum_{\inn} \frac{p_i p_j}{p \cdot \nb} =  \frac{1}{4Gi}\htilde^{(0) \out}_{i j}(-\nh) +  \sum_{\out}  \frac{p_i p_j}{p \cdot \nb}
\ee
where for convenience we defined
\be \label{defnb}
\nb^\mu := (1, -\nh).
\ee
Eq. \eqref{Tleadthm} is just  \eqref{leadthmTcovapp} evaluated at $-\nh$.

Consider now the log soft theorem
\begin{multline} \label{logthmTcovapp}
\frac{1}{4G}\htilde^{(\log)\out }_{i j}(\nh) + \sum_{\substack{ \out \\ m \neq 0}} \frac{p_{(i}  J^{\rad + }_{j)\rho} n^{\rho}}{p_i \cdot n} =  
\\ \frac{1}{4G}\htilde^{(\log)\inn }_{i j}(\nh) -  \sum_{\substack{\inn \\ m \neq 0}} \frac{p_{(i}  J^{\rad -}_{j)\rho} n^{\rho}}{p \cdot n}    + 4 G \bigg( P \cdot n \Big(\frac{1}{4Gi}\htilde^{(0) \inn}_{i j}(\nh) - \sum_{\inn }    \frac{p_{i}  p_{j} }{p \cdot n}  \Big) -  P_{i}P_{j} \bigg),
\end{multline}
where again, we have grouped incoming and outgoing terms and suppressed the particle index. Note that the ``extra'' term in the second line can alternatively be written in terms of outgoing data by using \eqref{leadthmTcovapp}. This is the key observation that allows to establish the time-reversal covariance of \eqref{logthmTcovapp}. 

To evaluate the time-reversal  of \eqref{logthmTcovapp} we  still need to specify the transformation properties of the log deviation vector under $\T$.  One can show (either from its  expression in terms of momenta, or through its relation with the asymptotic metric at timelike infinity) that the future and past radiative frame log deviation vectors are mapped into each other according to
  \be \label{Tc}
\T c_{\rad + , \out }^{\mu} = \left( -c_{\rad - , \inn }^{0}, \vec{c}_{\rad - , \inn }\right),
\ee
with analogue expression for $\T c_{\rad - , \inn }^{\mu}$. In particular, the angular momentum terms in \eqref{logthmTcovapp} transform as
\be \label{TJrad}
\T J^{\rad + }_{j \rho} n^{\rho} = - J^{\rad - }_{j \rho} \nb^{\rho}, \quad  \T J^{\rad - }_{j \rho} \nb^{\rho} = - J^{\rad + }_{j \rho} \nb^{\rho},
\ee
where we have kept implicit the in/out labels. Using \eqref{Tp}, \eqref{Thlog},  \eqref{TJrad} and, importantly, \eqref{leadthmTcovapp}, the time-reversal of \eqref{logthmTcovapp} takes the form
\begin{multline} \label{TlogthmTcovapp}
\T \eqref{logthmTcovapp} : \quad - \frac{1}{4G}\htilde^{(\log)\inn }_{i j}(-\nh) + \sum_{\substack{ \inn \\ m \neq 0}} \frac{p_{(i}  J^{\rad - }_{j)\rho} \nb^{\rho}}{p_i \cdot \nb} =  
\\ -\frac{1}{4G}\htilde^{(\log)\out }_{i j}(-\nh) -  \sum_{\substack{\out \\ m \neq 0}} \frac{p_{(i}  J^{\rad +}_{j)\rho} \nb^{\rho}}{p \cdot \nb}    + 4 G \bigg( P \cdot \nb \Big(\frac{1}{4Gi}\htilde^{(0) \inn}_{i j}(-\nh) - \sum_{\inn }    \frac{p_{i}  p_{j} }{p \cdot \nb}  \Big) -  P_{i}P_{j} \bigg).
\end{multline}
This relation is equivalent to \eqref{logthmTcovapp} evaluated at $-\nh$. We recall  that the appearance of the antipodal map is due to our conventions in \eqref{defhin}, see Footnote \ref{hinconvfnote} for related comments.

\section{Hyperboloid geometry formulas} \label{identitiesapp}

\subsection*{Timelike infinity} 

\begin{itemize}
\item Nontrivial Christoffel symbols of the metric \eqref{ds2timeH} (without considering those associated to the 2-sphere):
\be
 \Gamma^\rho_{\rho\rho}=-\frac{\rho}{1+\rho^2},  \quad  \Gamma^\rho_{A B}= -\rho (1+\rho^2) q_{AB}, \quad \Gamma^A_{B \rho}= \frac{1}{\rho}\delta^A_B
\ee

\item Commutator of covariant derivatives:
\be
[D_a,D_b] \w_c = \w_a h_{bc} -\w_b  h_{ac}.
\ee

\item Contractions of derivatives of the unit vector \eqref{Vxtime}
\be
D_aV^{\mu} D_b V_{\mu}=h_{ab}, \quad  D^a V^{\mu}D_a V^{\nu}=\eta^{\mu\nu}+V^{\mu}V^{\nu}.
\ee

\item Second derivative identities of the unit vector
\be
    (D_aD_b-h_{ab})V^{\mu}=0 \implies  (D^2-3)V^{\mu}=0.
\ee

\end{itemize}

\subsection*{Spatial infinity}

\begin{itemize}
\item Nontrivial Christoffel symbols of the metric \eqref{ds2spiH} (without considering those associated to the 2-sphere):
\be
 \Gamma^\t_{\t\t}=-\frac{\t}{1+\t^2},  \quad  \Gamma^\t_{A B}= \t (1+\t^2) q_{AB}, \quad \Gamma^A_{B \t}= \frac{\t}{1+\t^2}\delta^A_B
\ee

\item Commutator of covariant derivatives:
\be
[D_a,D_b] \w_c = - \w_a h_{bc} +\w_b  h_{ac}.
\ee

\item Contractions of derivatives of the unit vector \eqref{Vxspatial}
\be
D_aV^{\mu} D_b V_{\mu}=h_{ab}, \quad  D^a V^{\mu}D_a V^{\nu}=\eta^{\mu\nu}-V^{\mu}V^{\nu}.
\ee

\item Second derivative identities of the unit vector
\be
    (D_aD_b+h_{ab})V^{\mu}=0 \implies  (D^2+3)V^{\mu}=0.
\ee

\end{itemize}

\section{Sourced Beig-Schmidt expansion at timelike infinity} \label{BStimeapp}

In this appendix we extend the BS expansion at timelike infinity of \cite{cgw}, by explicitly including the matter stress tensor contribution. 
As in section \ref{timesec}, we treat simultaneously the $\t \to \pm \infty$ cases, omitting $\pm$ labels.

\subsection*{Asymptotic trajectories}
We start by describing the particles' asymptotic trajectories in BS coordinates, as they are the main input for the construction of the asymptotic stress tensor. In the notation of section \ref{prelsec},  the particles' trajectories in asymptotic Cartesian coordinates take the form
\begin{equation}
    X^{\mu}_i(s_i)  \stackrel{s_i \to \pm \infty}{=}   s_i V_i^{\mu}+\log |s_i| c_i^{\mu}+\cdots,
\end{equation}
with $s_i$ the  proper time of the $i$-th  particle. In BS coordinates  $(\t,x^a)$ the asymptotic geodesics take the form 

\begin{equation} \label{BSgeodesic}
    \tau_i (s_i)=s_i+O(\log |s_i|), \quad    x^a_i(s_i)=x^a_i+\frac{\log |s_i|}{s_i}c_i^a+\cdots,
\end{equation}
where  $ x_i^a$ (without the argument) is the hyperboloid coordinate  associated to $V^\mu_i$ and
\be
c^a_i  =  D^a V_\mu(x) c^\mu(x)|_{x=x_i} .
\ee

By studying the asymptotic geodesic equation \cite{sahoosen,gianni1} one can recover $c^a_i  = D^a \sigma(x)|_{x=x_i}$, in accordance with \eqref{cpmitosigma}.

\subsection*{Stress tensor}

The stress tensor due to  outgoing/incoming massive particles is 
\begin{equation} \label{gralTmunu}
    T^{\mu\nu}(X)=\sum_{i \in \pm }   m_i \int ds_i \frac{\delta^{(4)}(X-X_i(s_i))}{\sqrt{-g(X)}}  \frac{dX^{\mu}_i}{ds_i} \frac{dX^{\nu}_i}{ds_i},
\end{equation}
where $\mu, \nu$ stand for arbitrary spacetime coordinates, not necessarily Cartesian. Here and in the equations that follow we keep implicit the condition $m_i \neq 0$ in the sums.

We are interested in evaluating \eqref{gralTmunu} in BS coordinates for large $|\t|$.  Substituting \eqref{BSgeodesic} in \eqref{gralTmunu}, the various terms can be expanded according to
\be \label{del4del3}
\delta^{(4)}(X-X_i(s_i))= \delta(\tau-\tau_i(s_i))\Big(\delta^{(3)}(x,x_i)- \frac{\log |s_i|}{s_i}c_i^a\partial_a\delta^{(3)}(x,x_i)+O(s_i^{-1})\Big),
\ee
\be
\frac{1}{\sqrt{-g}}=|\tau|^{-3}\frac{1}{\sqrt{h}}\Big(1+\frac{2\sigma}{\tau}+O(\tau^{-2}\log|\tau|)\Big),
\ee
\be
\frac{dx_i^a(s_i)}{ds_i}=-c_i^a \frac{\log |s_i|}{s_i^2}+O(s_i^{-2}),
\ee
\be
\frac{d\t_i}{d s_i}=1+O(s_i^{-1}).
\ee

Performing the $s_i$ integral leads to
\begin{equation}
    T_{\tau\tau}=\frac{1}{\tau^3} T_{\tau\tau}^{(1)}+ \frac{\log|\tau|}{\tau^4} T_{\tau\tau}^{(\log)}+O(\tau^{-4}),
\end{equation}
\begin{equation}
    T_{\tau a}= \frac{\log|\tau|}{\tau^3} T_{\tau a}^{(\log)}+O(\tau^{-3}),
\end{equation}
\begin{equation}
    T_{ab}=  O(\t^{-3} \log^2 |\t|) ,
\end{equation}
with\footnote{In the rest of the paper we keep implicit the  factors of $\sqrt{h}$.}
\begin{equation}
     T_{\tau \tau}^{(1)}=\pm \sum_{i \in \pm } m_i \frac{\delta^{(3)}(x,x_i)}{\sqrt{h}} \equiv \rho(x)
\end{equation}
\begin{equation}
    T_{\tau a}^{(\log)}= \pm \sum_{i\in \pm} m_i \frac{\delta^{(3)}(x,x_i)}{\sqrt{h}}  c_{ia}=\rho \, c_{a}(x) \equiv j_a(x)
\end{equation}
\begin{equation}
    T_{\tau \tau}^{(\log)}=\mp \sum_{i \in \pm} m_iD_a\Big(\frac{\delta^{(3)}(x,x_i)}{\sqrt{h}} \Big) c_i^{a} =- D_a \Big( \rho \, c^a(x)\Big) =- D\cdot j  (x).
\end{equation}

\subsection*{Einstein tensor}

The Einstein tensor for the  Beig-Schmidt metric \eqref{BSgaugepm}, \eqref{gpmab} has the asymptotic form

\begin{equation}
    G_{\tau\tau}=\frac{1}{\tau^3} G_{\tau\tau}^{(1)}+ \frac{\log|\tau|}{\tau^4} G_{\tau\tau}^{(\log)}+O(\tau^{-4}),
\end{equation}
\begin{equation}
    G_{\tau a}=\frac{1}{\tau^2} G_{\tau a}^{(1)}+ \frac{\log|\tau|}{\tau^3} G_{\tau a}^{(\log)}+O(\tau^{-3}),
\end{equation}
\begin{equation}
    G_{ab}=\frac{1}{\tau} G_{ab}^{(1)}+ \frac{\log|\tau|}{\tau^2} G_{ab}^{(\log)}+O(\tau^{-2}),
\end{equation}
with
\begin{equation}
    G^{(1)}_{\tau\tau}=2(D^2-3)\sigma, \quad G_{\tau a}^{(1)}=-\frac{1}{2} D^bk_{ab}, 
\end{equation}
\be
G_{a b}^{(1)}= -\frac{1}{2} (D^2+3)k_{ab}+D_{(a}D^ck_{b)c}-\frac{1}{2}D^cD^dk_{cd}\, h_{ab},
\ee
\be
    G_{\tau\tau}^{(\log)} = -i-\frac{1}{2}D^2i+\frac{1}{2}D^aD^bi_{ab}, \quad    G_{\tau a}^{(\log)} =   D_a i- D^bi_{ab},
\ee
\begin{equation}
    G_{ab}^{(\log)} = -i_{ab} -\frac{1}{2}D^2 i_{ab} -\frac{1}{2} D_aD_b i + D_{(a}D^ci_{b)c}+ \frac{1}{2} D^2i \, h_{ab}-\frac{1}{2}D^cD^di_{cd}\, h_{ab}.
\end{equation}
where
\be
i= h^{ab} i_{ab}.
\ee

\subsection*{Einstein equations}
Equating the asymptotic  Einstein tensor to ($8 \pi G$ times) the asymptotic stress tensor, one finds:
\begin{itemize}
\item Equations for $k_{ab}$:
\be \label{eck}
G_{\tau a}^{(1)}= G_{a b}^{(1)}=0.
\ee
\item
 Equation for $\sigma$:
\be\label{eqsig}
(D^2-3)\sigma=4\pi G\rho.
\ee
\item Equations for $i_{ab}$:
\begin{equation}\label{i}
 i+\frac{1}{2}D^2 i -\frac{1}{2}D^aD^b i_{ab}= 8\pi G D\cdot j,
\end{equation}
\begin{equation}\label{ia}
  D_ai-D^bi_{ab}=8\pi G j_a,
\end{equation}
\begin{equation}\label{iab}
 i_{ab}+\frac{1}{2}D^2 i_{ab}+\frac{1}{2}D_a D_b i -D_{(a}D^ci_{b)c} -\frac{1}{2}D^2 i \, h_{ab}+\frac{1}{2}D^cD^di_{cd}\, h_{ab}=0.
\end{equation}
\end{itemize}

Equations \eqref{eck} are automatically solved by the ``pure gauge'' form of $k_{ab}$ given in Eq. \eqref{kabitoPhitime} \cite{dehouck,cgw}.
The solution to \eqref{eqsig} is well-known and will be reviewed in the next section.  
We now discuss how to simplify the   system of equations  \eqref{i}, \eqref{ia}, \eqref{iab} for $i_{ab}$. 

Taking $1/2$ of the divergence of \eqref{ia} and subtracting  the result to \eqref{i} leads to 
\begin{equation}\label{eqi1}
    i=4\pi G D\cdot j.
\end{equation}
Using \eqref{eqi1}, Eq. \eqref{ia} becomes
\begin{equation}\label{eqi2}
    D^b i_{ab}=4\pi G \Big( D_a D\cdot j -2  j_a\Big).
\end{equation}
Substituting \eqref{eqi1} and \eqref{eqi2} in \eqref{iab} leads to
\begin{equation}\label{eqi}
    (D^2+2)i_{ab}=4\pi G \Big((D_aD_b+2h_{ab})D\cdot j-\frac{1}{2} D_{(a}j_{b)}\Big).
\end{equation}
Finally, using \eqref{eqi1} to extract the traces in \eqref{eqi2} and \eqref{eqi} leads to \eqref{iabecstime}. We note that when $j_a=0$ the resulting set of equations for $i_{ab}$ reduce to the ones presented in \cite{cgw}.

\section{Green's functions}

In this appendix we construct the Green's functions that determine the required metric coefficients at timelike and spatial infinity.\footnote{See \cite{eyhe,coito,Briceno:2025ivl} for similar analysis in the context of electromagnetism and massless scalars;  \cite{cedric,henntrem,henntrscalar,comprobds3,compererobert} for alternative treatments based on spherical harmonic decomposition; and  \cite{deBoer:2003vf,Pasterski:2017kqt,Donnay:2018neh,Pasterski:2020pdk} for related constructions in the context of flat-space holography. } 

\subsection{Timelike infinity} \label{timegreenapp}

We start by reviewing the construction of the Green's function in \eqref{sigmaitogreentime} (see e.g. \cite{chipum2}), and then apply the same ideas for  \eqref{iabitoj}.

\subsubsection*{Green's function for $\sigma$}
We want to find $\G(x,x') $ such that
\be \label{d2minus3G}
(D^2-3) \G(x,x') =  \delta(x,x').
\ee
Our basic assumption is that the Green's function is symmetric under exchange of $x$ and $x'$ so that\footnote{The geodesic distance between $x$ and $x'$ is given by $d(x,x')= \cosh^{-1}\chi$.}
\be   \label{defchitime}
\G(x,x') = g(\chi), \quad   \chi := - V^\mu V'_\mu,
\ee
where  $V^\mu $ and ${V'}^\mu$ are the unit vectors associated to the points $x$ and $x'$ according to  Eq. \eqref{Vxtime}.  
Using the identities given in appendix \ref{identitiesapp} one has
\be \label{d2min3gtime}
(D^2-3) g =    (\chi^2 -1) \ddot g +3 \chi \dot g  - 3 g , 
\ee
where the dot in \eqref{d2min3gtime} denotes derivative with respect to $\chi$. The vanishing of \eqref{d2min3gtime} for  $\chi \neq 1$ (i.e. $x \neq x')$ leads to a second order differential equation, whose general solution is
\be \label{gsigmagreentime}
g = A  \left( \frac{2\chi^2-1}{\sqrt{\chi^2-1}}   + B \chi \right),
\ee
with $A$, $B$ integration constants.  The constant $A$ is fixed by requiring   compatibility with the  Dirac delta in \eqref{d2minus3G} when  $\chi \to 1$ \cite{chipum2} (see the last subsection of this appendix for similar considerations in the context of $i_{ab}$) from where one gets
\be
A=  - \frac{1}{4\pi}.
\ee
The value of $B$ depends on the choice of log translation frame. In the radiative frame we are interested in, we have
\be \label{choiceBradframegtime}
g \stackrel{\chi \to \infty}{\to} 0 \implies B=-2.
\ee
The resulting Green's function is then
\be \label{greensigmatime}
\G(x,x') =  - \frac{1}{4\pi} \left(   \frac{2\chi^2-1}{\sqrt{\chi^2-1}} -2 \chi\right) .
\ee

We finally study the large $\rho$ behaviour of \eqref{greensigmatime}. In this limit we have 
\be \label{largerhochi}
\chi  \stackrel{\rho \to \infty}{=} \rho \, \psi + O(1/\rho), \quad \psi:=  - n^\mu V'_\mu.
\ee
Substituting \eqref{largerhochi} in \eqref{greensigmatime} we get
\be 
\G(x,x')  \stackrel{\rho \to \infty}{=}  -\frac{1}{16 \pi \rho^3}  \frac{1}{\psi^3} + \cdots,
\ee
which corresponds to Eq. \eqref{asymgreensigmatime}.

\subsubsection*{Green's function for $i_{\langle ab\rangle}$}

We now turn to the problem given by the last two equations in \eqref{iabecstime}, 
\be \label{pdeiabtimeapp}
\begin{aligned}
  D^b  i_{\langle ab \rangle} &=8\pi G \left(   \tfrac{1}{3}D_a D \cdot j- j_a \right) , \\
(D^2+2)i_{\langle ab \rangle} &=4\pi G\left(  D_{\langle a} D_{b \rangle} D \cdot j    - 4 D_{\langle a}j_{b\rangle}\right).
\end{aligned}
\ee
Our starting point is to consider an ansatz of the form
\be \label{iabeqdaib}
i_{\langle ab \rangle}= D_{\langle a}  i_{b \rangle} ,
\ee
where $i_a$ is a vector field on $\H^\pm$. It is a priori not obvious that the most general solution to \eqref{pdeiabtimeapp} can be written in this form, but we shall later provide an argument as to why this must be the case. In order for \eqref{iabeqdaib} to satisfy \eqref{pdeiabtimeapp}, the vector field has to obey\footnote{As in the original set of equations \eqref{pdeiabtimeapp}, one can verify the compatibility of Eqs. \eqref{pdeiatimeapp}  by comparing  the Laplacian of the first one  with the divergence of the second one.}
\be
\begin{aligned} \label{pdeiatimeapp}
  D^a  i_{a} &= 4\pi G D\cdot j \\
(D^2-2)i_{a} &= 8\pi G\Big( \frac{1}{2}D_{a}  D\cdot j -2 j_{a}\Big) .
\end{aligned}
\ee
We will first construct the Green's function for \eqref{pdeiatimeapp}, and then use it to obtain the Green's function for the original problem \eqref{pdeiabtimeapp}.

\subsubsection*{Green's function for \eqref{pdeiatimeapp}}
Let us write the solution to \eqref{pdeiatimeapp} as
\be \label{iaintd3ypGF}
i_a(x) = 8 \pi G \int d^3 x' \G_{a}^{b'}(x,x') j_{b'}(x'),
\ee
where  $\G_{a}^{b'}(x,x')$ is defined by the conditions
\be
\begin{aligned} \label{eqsgreeniatime}
  D^a  \G_{a}^{b'}(x,x') &= - \tfrac{1}{2} D^{b'} \delta(x,x'),  \\
(D^2-2)\G_{a}^{b'}(x,x') &= -\big( \tfrac{1}{2}D_{a}D^{b'} +2 \delta^{b'}_{a} \big)\delta(x,x').
\end{aligned}
\ee

By Lorentz invariance, it should be possible to express $\G_{a}^{b'}$ in terms of $\chi \equiv - V \cdot V'$ and its derivatives. The most general form that is compatible with the tensorial structure is
\be \label{Gbc}
\G_{a}^{b'}(x,x') = f(\chi) D_{a} \chi   D^{b'} \chi + g(\chi)  D_{a}  D^{b'} \chi,
\ee
with $f$ and $g$ free functions. Using the identities given in appendix \ref{identitiesapp} one finds 
\be  \label{divGiaapp}
D^a \G_{a}^{b'} = \left[ (\chi^2-1) \dot f  + 4 \chi f  + \chi \dot g  + 3 g \right] D^{b'} \chi 
\ee
\begin{multline}  \label{lapGiaapp}
(D^2-2) \G_{a}^{b'}  = \left[ (\chi^2-1) \ddot f  + 7\chi  \dot f  +2 f + 2 \dot g \right] D_a \chi D^{b'} \chi  \\
+ \left[ (\chi^2-1) \ddot g  + 3 \chi \dot g  - g + 2 \chi f   \right]D_a  D^{b'} \chi . 
\end{multline}
For $x\neq x'$ ($\chi \neq 1$) the coefficients in square brackets in \eqref{divGiaapp} and \eqref{lapGiaapp}  should vanish. These yields three self-consistent equations  that can be used to determine $f$ and $g$. The most general solution  is 
\be
\begin{aligned} \label{gralsolfg}
f(\chi)& = A \left( \frac{\chi \left(4 \chi^4-10 \chi^2+9\right)}{3 \left(\chi^2-1\right)^{5/2}} +B \right) \\
g(\chi)& = -A \left( \frac{4 \chi^4-6 \chi^2+3}{3 \left(\chi^2-1\right)^{3/2}}+ B \chi \right) ,
\end{aligned}
\ee
where $A$ and $B$ are integration constants.  As in the case of $\sigma$, the constant $A$ is  fixed by ensuring  the  singularity of the Green's function at $\chi = 1$ is in accordance with the (derivatives of) Dirac deltas in \eqref{eqsgreeniatime}. This leads to (see the end of this appendix for a proof)
\be \label{Agreeniabspi}
A= \frac{3}{8 \pi} .
\ee

The dependence on the  constant $B$ disappears upon taking the  symmetrized derivative \eqref{iabeqdaib}, and thus plays no role for the Green's function of $i_{ab}$.\footnote{The ambiguity associated to the constant $B$ corresponds to the possibility of adding a term proportional to a global rotation $i_a \to i_a + L^{\mu \nu} V_{[\mu} D_a V_{\nu]}$ with constant $L^{\mu \nu}$. The situation is similar to the ambiguity of  $B$ in \eqref{gsigmagreentime}, although here we lack an interpretation in terms of residual diffeomorphisms. \label{logrotationfnote}}  From the point of view of $i_a$, however, it is natural to fix it by requiring decaying boundary condition. We will return to this point later on.

\subsubsection*{Green's function for \eqref{pdeiabtimeapp}}
Let us write the solution to \eqref{pdeiabtimeapp} as
\be \label{iabgreentimeapp}
i_{\langle ab \rangle}(x) = 8 \pi G \int d^3 x' \G_{ab}^{c'}(x,x') j_{c'}(x').
\ee
As before, we assume the Green's function is to be constructed out of $\chi$ and its derivatives. The most general expression compatible with the tensorial structure is
\be \label{greeniabansatztime}
 \G_{ab}^{c'}(x,x')  = F(\chi) D_{\langle a} \chi D_{b \rangle} \chi D^{c'} \chi + G(\chi) D_{\langle a} \chi D_{b \rangle}  D^{c'} \chi.
\ee

Let us now compare this expression with what one gets from \eqref{Gbc} and \eqref{iabeqdaib}: 
\be 
 D_{\langle a } \G_{b \rangle}^{c'}(x,x') =  \dot f(\chi) D_{\langle a} \chi D_{b \rangle} \chi D^{c'} \chi +\big[f(\chi)+ \dot g(\chi) \big]D_{\langle a} \chi D_{b \rangle}  D^{c'} \chi. \label{DaGbgreenapp}
\ee

We see that \eqref{greeniabansatztime} can always be brought into the form  \eqref{DaGbgreenapp}, provided we do the identification, 
\be \label{FGitofg}
\begin{aligned}
F(\chi) &= \dot f(\chi), \\
G(\chi) &=  f(\chi)+ \dot g(\chi).
\end{aligned}
\ee

So far we are not imposing any restrictions on $F$ and $G$: Given any pair $(F,G)$ we can always find  $(f,g)$ satisfying \eqref{FGitofg}, so that \eqref{greeniabansatztime} is given by the  total derivative expression \eqref{DaGbgreenapp}.  The argument shows that the solution to \eqref{pdeiabtimeapp} is indeed of the form \eqref{iabeqdaib}. It then follows that the Green's function \eqref{iabgreentimeapp} is given by \eqref{DaGbgreenapp}, with $f$ and $g$ given by \eqref{gralsolfg}, \eqref{Agreeniabspi}. As expected, the dependance on the constant $B$ disappears in \eqref{FGitofg}.

\subsubsection*{Asymptotic behavior}
We finally study the $\rho \to \infty$ behavior of the Green's functions \eqref{iaintd3ypGF} and \eqref{iabgreentimeapp}. We first focus on  $i_a$.

For large enough $\rho$ we can set to zero the source terms in \eqref{pdeiatimeapp}. The resulting asymptotic equations admit only two possible asymptotic behaviors on $i_A$,
\be \label{iAOrhopm2}
i_A \stackrel{\rho \to \infty}{=}  O(\rho^{\pm 2}). 
\ee
 The plus case corresponds to homogenous solutions to \eqref{pdeiatimeapp} and represent superrotation vectors fields $i^a=h^{ab}i_b$ on $\H^\pm$, as defined in \cite{clmassive}. The minus case corresponds to solutions with source terms in the interior; these are the ones we are interested in. Let us explicitly see how they arise from the Green's function. 

Consider first the large $\chi$ expansion of \eqref{gralsolfg}
\bal \label{fgchitoinf}
f(\chi)& \stackrel{\chi \to \infty}{=} A \left( (4/3+B) +\frac{1}{2 \chi^4}+O(1/\chi^{6}) \right) \\
g(\chi)&  \stackrel{\chi \to \infty}{=} -A \left( (4/3+B) \chi+\frac{1}{2 \chi^3}+O(1/\chi^{5})\right).
\eal
Recalling that $O(\chi)=O(\rho)$ one can check that, indeed, these produces $ O(\rho^{\pm 2})$ terms in $i_A$, consistent with \eqref{iAOrhopm2}. Furthermore, the  $O(\rho^2)$ part can be eliminated  by setting\footnote{This is analogous to condition \eqref{choiceBradframegtime}, see  footnote \ref{logrotationfnote}. We recall that $i_{\langle ab \rangle}$ is independent of $B$.}
\be \label{valueBiabapp}
B=-4/3.
\ee

Using \eqref{fgchitoinf} and \eqref{valueBiabapp} in \eqref{Gbc} along with \eqref{largerhochi}
\be
\chi  \stackrel{\rho \to \infty}{=} \rho \, \psi + O(1/\rho), \quad \psi \equiv - n \cdot V'
\ee
we get
\be \label{Gabasymapp}
\G_{A}^{b'}(x,x')  \stackrel{\rho \to \infty}{=}-\frac{A}{2 \rho^2 \psi^4}  \ell^{b'}_A   + O(\rho^{-4})
\ee
where 
\ba \label{defellapptime}
\ell^{b'}_A & :=& -  D_A \psi D^{b'} \psi +\psi D_A D^{b'}\psi \\
&= & \partial_A n^{\mu} n^{\nu} D^{b'} V'_{[\mu} V'_{\nu]} .
\ea

A short computation shows that, for a vector field with the above fall-offs one has\footnote{The $i_\rho$ component decays as $1/\rho^5$ and does not contribute to the leading term  of $D_{\langle \rho } i_{A \rangle} $.}
\be \label{irhAfallitoiA}
i_A \stackrel{\rho \to \infty}{=} \frac{1}{\rho^{2}} \ov{0}{i}_A + \cdots \implies D_{\langle \rho } i_{A \rangle} \stackrel{\rho \to \infty}{=} -\frac{2}{\rho^{3}} \ov{0}{i}_A.
\ee
Applying \eqref{irhAfallitoiA} to \eqref{Gabasymapp} we obtain,
\be \label{Grhoabasymapp}
\G_{\rho A}^{b'}(x,x')  \stackrel{\rho \to \infty}{=}\frac{A}{ \rho^3 \psi^4} \ell^{b'}_A +  O(\rho^{-5}),
\ee
which corresponds to \eqref{asymGrhoAbp} upon using \eqref{Agreeniabspi}.

\subsubsection*{Eq. \eqref{Agreeniabspi}}
In this final subsection, we determine the constant $A$ in \eqref{gralsolfg}. The idea is that in the limit where $x$ is close to $x'$, we can approximate \eqref{eqsgreeniatime} by their flat space counterparts, where they can easily be integrated. Let us fix
\be
{V'}^\mu=(1,\vec{0}) \implies \chi = \sqrt{1+\rho^2},
\ee
and define  local Cartesian coordinates by,
\be \label{defloccartcoord}
\xb^i :=  \rho\, \nh^i ,
\ee
in terms of which, the hyperboloid metric takes the form
\be
h_{ij}=\delta_{ij}- \frac{1}{\rho^2+1}\xb_i \xb_j,
\ee
where $\xb_i =\xb^i$ and $\rho^2 = \xb^i \xb_i$.  

We will focus on the second of Eqs. \eqref{eqsgreeniatime}. In the $\rho \to 0$ limit and in the coordinates \eqref{defloccartcoord}, the equation at $x'=0$ takes the form
\be \label{lapcartesiantimeapp}
\left(\partial^k \partial_k  + O(\rho^0) \right) \G_{i}^{j'}(\xb,0) =  -\left( \tfrac{1}{2}\partial_{i} \partial^{j'} +2 \delta^{j'}_{i} \right) \left.  \delta(\xb,\xb') \right|_{\xb'=0}
\ee
where $\partial^i=\partial_i = \partial/\partial \xb^i$. We note that the $O(\rho^0)$ piece in \eqref{lapcartesiantimeapp} includes  multiplicative terms as well as terms with one derivative, of the form $O(\rho) \partial_k$ since each derivative counts as   $O(\rho^{-1})$. 

Using the standard flat-space relation
\be
\delta(\xb,0)=  - \frac{1}{4\pi} \partial^k \partial_k  \left. \frac{1}{| \xb-\xb'|} \right|_{\xb'=0},
\ee
the solution to   \eqref{lapcartesiantimeapp} is found to be
\ba
\G^{ j'}_{i}(\xb,0) &= & \frac{1}{4\pi}  \left( \tfrac{1}{2}\partial_{i} \partial^{j'} +2 \delta^{j'}_{i} \right) \left. \frac{1}{| \xb-\xb'|}  \right|_{\xb'=0} + O(\rho^{-1})\\
&= &  \frac{1}{4\pi} \left(  \frac{1}{2} \frac{\delta_{i}^{j'}}{\rho^3} - \frac{3}{2} \frac{\xb_i \xb^{j'}}{\rho^5} \right)  + O(\rho^{-1}). \label{flatgreen}
\ea

Let us now consider the  $\rho \to 0$ expansion of  \eqref{Gbc}. The derivatives of $\chi$ in the coordinates \eqref{defloccartcoord} at $x'=0$ take the form\footnote{To evaluate the  primed derivatives one needs the expression for $\chi$ away from $x'=0$, $\chi=\sqrt{1+\rho^2}\sqrt{1+{\rho'}^2}-\xb^k \xb_k$.}
\bal \label{cartderchitime}
 \partial'_{i}\chi &=-\xb_{i},  \quad  \partial_i \partial'_{j}\chi =-\delta_{ij}, \\
\quad \partial_i \chi &= \frac{\xb_i}{\chi} = \xb_i (1- \rho^2/2+O(\rho^4)),
\eal
while the short distance expansion of the functions \eqref{gralsolfg} read
\bal \label{rhotozerofgiabtime}
f(\chi) & \stackrel{\rho \to 0}{=} A\left( \frac{1}{\rho^5}-\frac{1}{6}\frac{1}{\rho^3} + \cdots \right),\\
g(\chi) & \stackrel{\rho \to 0}{=} -\frac{A}{3} \left(\frac{1}{\rho^3}+2 \frac{1}{\rho}+ \cdots \right).
\eal
Substituting  \eqref{cartderchitime},  \eqref{rhotozerofgiabtime} in \eqref{Gbc} yields
\be
\G^{ j'}_{i}(\xb,0)  \stackrel{\rho \to 0}{=} A \left( -\frac{1}{\rho^5} \xb_i \xb^{j'} + \frac{1}{3}\frac{1}{\rho^3 } \delta_{i}^{j'}  \right) +O(\rho^{-1}).
\ee
By comparing with \eqref{flatgreen}, we conclude that  $A=3/(8 \pi)$.

\subsection{Spatial infinity}\label{greenspiapp}

In this subsection we construct   Green's function for $\sigma$ and $i_{ab}$ that solve their respective equations in terms of asymptotic final or initial data.

\subsubsection*{Green's function for $\sigma$}

Let us for concreteness   focus on the future radiative frame  solution. We want to find $\sigma^{\rad +}$ such that
\be \label{sigmaprobspi}
(D^2 +3 ) \sigma^{\rad +} =0, \quad  \sigma^{\rad +}(\t,\phi) \stackrel{\t \to + \infty}{=} \sigmazero^+(\phi)/\t^3 + \cdots,
\ee
for a given final data $\sigmazero^+(\phi)$. This problem may be solved by first considering the finite-$\t$ final value problem to the PDE (e.g. through a  3d ``bulk-to-bulk'' advanced Green's function) and then take $\t \to \infty$.\footnote{See \cite{eyhe} for an implementation of this approach in the context of electrodynamics.} It is however simpler to find directly  the ``boundary-to-bulk'' Green's function 
\be \label{defGplusspi}
(D^2 +3 ) G(x,\phi')  =0, \quad G(\t,\phi,\phi') \stackrel{\t \to + \infty}{=} \delta(\phi,\phi')/\t^3 + \cdots,
\ee
in terms of which the solution to \eqref{sigmaprobspi} reads
\be \label{defGreenplussigma}
 \sigma^{\rad +}(x) = \int d^2 \phi' G(x,\phi') \sigmazero^+(\phi') .
\ee

To solve for \eqref{defGplusspi} we consider the ansatz
\be \label{Gitogspisigma}
G(x,\phi')  = g(\psi), \quad \psi := V^\mu(x) n_\mu(\phi'),
\ee
that ensures Lorentz covariance.  From the identities reviewed in appendix \ref{identitiesapp}, $g$ should satisfy 
\be \label{D2greenspi}
(D^2 +3 ) g = - \psi^2 \ddot g -3 \psi \dot g +3 g =0,
\ee
where the dot denotes derivative with respect to $\psi$. The general \emph{non-distributional} solution to this equation is $g = A \psi + B/\psi^3$, with $A$, $B$ integration constants. The first term leads to a  pure log translation while the second term leads to a  Green's function with boundary value $1/\psi^3 \stackrel{\t \to + \infty}{\sim} \t \delta(\phi,\phi')$.\footnote{Relevant for solving $\Phi(x)$ in \eqref{kabitoPhispi} in terms of the supertranslation Goldstone $C(\nh)$ \cite{clmassive,phigoldstone,cedric}.} To find solutions  with the desired decaying behavior one needs to consider distributional solutions. These can be obtained  by doing the replacement $\psi \to \psi + i 0$ in the regular  solutions \cite{gelfandshilov}, leading to $g = C \psi \theta(\psi) + D \ddot \delta (\psi)$ where $\theta$ is the step function, $\ddot \delta$ the second derivative of the Dirac delta, and  $C$, $D$ new integration constants.\footnote{Ec. \eqref{D2greenspi} is a particular case of  $(D^2+(n^2-1) ) g=0, n \in \mathbb{N}^+$, with general solution  $g=A \psi^{n-1}+B/\psi^{n+1}+ C\psi^{n-1}\theta(\psi) + D \delta^{(n)}(\psi)$. It would be interesting to understand the relationship with the scalar harmonics studied in \cite{comprobds3}, which are defined through the same family of equations.}  As we now discuss, the  desired boundary condition is obtained by setting $D=0$ and $C=4/\pi$.  

The asymptotic behavior of  $\psi$ at large positive $\t$ is
\be
\psi \stackrel{\t \to + \infty}{=} \t \, n(\phi) \cdot n(\phi') + \cdots.
\ee
One can show  (see e.g. appendix E of \cite{aview}) that in this limit, 
\be \label{psithetadelta}
\psi \, \theta(\psi) \stackrel{\t \to + \infty}{=}   \frac{\pi}{4 \t^3} \delta(\phi,\phi') + \cdots,
\ee
and therefore 
\be \label{gsigmaspi}
g(\psi)= \frac{4}{\pi} \psi \, \theta(\psi)
\ee
is the solution to  \eqref{D2greenspi} with the desired boundary value \eqref{defGplusspi}. 

In order to establish the matching properties between future and past, let us now evaluate  \eqref{gsigmaspi} at $\t \to -\infty$. The asymptotic behavior of  $\psi$ for large negative $\t$ is
\be \label{psiminif}
\psi \stackrel{\t \to - \infty}{=} \t \,  n(\A \phi) \cdot n(\phi') + \cdots
\ee
while the analogue of Eq. \eqref{psithetadelta} in this limit is
\be \label{psithetadeltam}
\psi \, \theta(-\psi) \stackrel{\t \to - \infty}{=}   \frac{\pi}{4 \t^3} \delta(\A \phi,\phi') + \cdots
\ee
Writing $ \theta(\psi)=1- \theta(-\psi)$ in \eqref{gsigmaspi} and using \eqref{psithetadeltam} we conclude that
\be \label{gmininf}
g(\psi) \stackrel{\t \to - \infty}{=}  \frac{4 \psi}{\pi}  -\frac{1}{\t^3} \delta(\A \phi,\phi') + \cdots.
\ee

In \eqref{gmininf} we purposely did not  expand the first term as in \eqref{psiminif} since it is, by itself, a solution of \eqref{D2greenspi}. This means that the remaining pieces must add up to a solution of  \eqref{D2greenspi}.  We now show that such remaining terms are the  potential in the past radiative log frame. To this end,  let us smear \eqref{gmininf} with $\sigmazero^+(\phi')$ as in \eqref{defGreenplussigma}.  Using \eqref{Pmuitosigmazero}, the integrated version of \eqref{gmininf} then reads
\be \label{sigmaradplusatmininf}
 \sigma^{\rad +}(x) \stackrel{\t \to - \infty}{=} 4 G P \cdot V(x)  -\frac{1}{\t^3} \sigmazero^+(\A \phi) + \cdots.
\ee

Comparing with \eqref{difradssigmaspi} we see that indeed the second term in the RHS of  \eqref{sigmaradplusatmininf}, with dots included, is  $ \sigma^{\rad -}(x)$, with  boundary value at $\t \to -\infty$ given by $\sigmazero^-(\phi) =-\sigmazero^+(\A \phi)$.

\subsubsection*{Green's function for $i_{ab}$}
As in the timelike case, the solution to \eqref{eomiabspi} can be written in terms of a ``vector potential'' $i_a$, such that
\be \label{iabitoiaspi}
i_{ab}= D_{\langle a}  i_{b \rangle} ,
\ee
with $i_a$ satisfying 
\be \label{PDEiaspi}
 D^a  i_{a} = 0, \quad (D^2+2)i_{a} =0.
\ee

For  large $\t$,  \eqref{PDEiaspi} implies  $i_A=O(\t^{\pm 2})$. We are interested in the decaying solutions (modulo a few growing solutions in the kernel of \eqref{iabitoiaspi}, see below).  Consider for concreteness final value conditions  
\be \label{iAlargtauspi}
i_{A}(\t,\phi) \stackrel{\t \to + \infty}{=}  \frac{\izero^+_{A}(\phi)}{\t^2} + \cdots.
\ee

The solution to \eqref{PDEiaspi} under \eqref{iAlargtauspi} can be written as
\be \label{iaitoIBspi}
i_a(x) = \int d^2 \phi' G_a^{B'} (x,\phi')   \izero^+_{B'}(\phi'),
\ee
with $G_a^{B'}$ satisfying
\ba
D^a G_a^{B'} &=&0,  \label{divGiabspi} \\
  (D^2+2) G_a^{B'} &=&0 , \label{waveGiabspi} \ \\
 G_A^{B'}(\t,\phi,\phi') &\stackrel{\t \to + \infty}{=} & \delta_A^{B'}\delta(\phi,\phi')/\t^2 + \cdots. \label{largetauGABspi}
\ea

As in \eqref{Gitogspisigma}, we consider an ansatz where the Green's function depends on $\psi$,
\be \label{GaBitogspi}
 G_a^{B'} (x,\phi')=\ell_a^{B'} g(\psi)
\ee
with
\be \label{defellspi}
 \ell_a^{B'}  := -  D_a \psi D^{B'} \psi +\psi D_a D^{B'}\psi 
\ee
capturing the tensorial structure.\footnote{One could consider the more general ansatz $f_1(\psi) D_a \psi D^{B'} \psi+ f_2(\psi) D_a D^{B'}\psi$. The field equations however restrict these two terms  to combine in the form \eqref{iaitoIBspi}. This is a spatial infinity  counterpart to the  timelike infinity expression \eqref{Gabasymapp}.} For later reference, we note that the large $\t$ limit of the sphere-sphere components of \eqref{defellspi} is
\be \label{largetauellspi}
 \ell_A^{B'} \stackrel{\t \to + \infty}{=} \t^2 \big(-(D_A n \cdot n') \, (n \cdot D^{B'} n') +   n \cdot n' D_A n \cdot D^{B'} n' \big)+ O(\t^0).
\ee
where $n' \equiv n(\phi')$.  

Using the identities of appendix \ref{identitiesapp}, one finds that the divergence-free condition \eqref{divGiabspi} is automatically satisfied by \eqref{GaBitogspi}, while   \eqref{waveGiabspi} translates into
\be \label{eqgiaspi}
\psi^2 \ddot g + 5 \psi \dot g =0.
\ee
The general non-distributional solution to \eqref{eqgiaspi} is  $g = A  + B/\psi^4$. The constant term represents a redundancy in the definition of $i_a$ that goes away when evaluating $i_{ab}$  in \eqref{iabitoiaspi}.\footnote{The same phenomenon occurs in the timelike infinity case, see footnote \ref{logrotationfnote}.} The second term can be shown to lead to a growing solution,  $i_{A}=O(\t^2)$, relevant for the boundary-to-bulk Green's function for superrotations. To get decaying boundary values  we need to consider   distributional solutions to \eqref{eqgiaspi},  given by   $g=   C \theta(\psi) + D \delta^{(3)}(\psi)$.  In both cases the large $\t$ limit is proportional to (possibly derivatives of)  $\delta(\phi,\phi')$, which implies the first term in \eqref{largetauellspi} is subdominant with respect to the second one.  Using this fact, along with \eqref{psithetadelta} and $D_A n(\phi) \cdot D^{B'} n(\phi')\big|_{\phi'=\phi}= \delta_A^{B'}$ we find
\be \label{ellthetapinf}
  \ell_A^{B'} \theta(\psi) \stackrel{\t \to + \infty}{=} \frac{\pi}{4 \t^2} \delta_A^{B'} \delta(\phi,\phi') + \cdots.
\ee
We then conclude that the solution to \eqref{eqgiaspi} that is consistent with the boundary value   \eqref{largetauGABspi} is given by
\be
g(\psi)= \frac{4}{\pi}  \theta(\psi).
\ee

Let us finally evaluate the  $\t \to -\infty$ limit of this Green's function.  The analogue of \eqref{ellthetapinf} in this limit is
\be 
  \ell_A^{B'} \theta(-\psi) \stackrel{\t \to - \infty}{=} \frac{\pi}{4 \t^2} \A_* \delta_A^{B'} \delta(\A \phi,\phi') + \cdots,
\ee
where $\A_* \delta_A^{B'} =D_A n(\A \phi) \cdot D^{B'} n(\phi')\big|_{\phi'=\A \phi} $ implements the pull-back map on the index $A$.

The large negative $\t$ limit of the sphere-sphere component of the Green's function can then be written as
\be \label{GABpminf}
G_A^{B'}(\t,\phi,\phi') \stackrel{\t \to - \infty}{=} \frac{4}{\pi}  \ell_A^{B'} - \frac{1}{ \t^2} \A_* \delta_A^{B'} \delta(\A \phi,\phi') + \cdots,
\ee
where similarly to what we did in Eq. \eqref{gmininf}, we kept unexpanded the first term. Using \eqref{GABpminf} in \eqref{iaitoIBspi} we get
\be \label{iAtauminusspi}
i_{A}(\t,\phi) \stackrel{\t \to - \infty}{=}  R_A(x)  -\frac{1}{\t^2} \A_*  \izero^+_{A}(\phi)  + \cdots.
\ee
where  $R_A$ is the sphere component of the Killing vector field
\be
R_a(x) = \frac{4}{\pi} \int d^2 \phi'  \ell_a^{B'} \, \izero^+_{B'}(\phi').
\ee
This term vanishes when evaluating the symmetrized derivative \eqref{iabitoiaspi}, and hence we conclude that $i_{ab}$ has decaying boundary conditions at both future and past infinities. Using \eqref{iAlargtauspi} and \eqref{iAtauminusspi} to evaluate $i_{\t A}=  D_{\langle \t}  i_{A \rangle} $ at these limits one finds\footnote{Similar to the timelike case, the $i_\t$ component decays as $1/\t^5$ and does not contribute to $\izero_{\t A}$.}
\be 
i_{\t A}(\t,\phi) \stackrel{\t \to \pm \infty}{=}  \frac{\izero^\pm_{\t A}(\phi)}{\t^3} + \cdots
\ee
with
\be \label{izerotauAitoizeroA}
\izero^+_{\t A}  =  - 2  \izero^+_{A} , \quad \quad \izero^-_{\t A} =   2 \A_* \izero^+_{A}.
\ee

\subsubsection*{Eq. \eqref{delLitauAplus}}

It is simpler to  work at the level of the vector $i_a$. Eq. \eqref{deliabspi} can be written as
\be
\delta_L i_{ab} = D_{\langle a }\delta_L i_{b \rangle}
\ee
with
\be \label{delLia}
\delta_L i_a = -2 l D_a \sigma + 3 \sigma D_a l + D^c l D_c D_a \sigma.
\ee
Evaluating the large $\t$ limit of \eqref{delLia} leads to
\be \label{largetaudelLiA}
\delta_L i_A(\t,\phi)  \stackrel{\t \to + \infty}{=} \frac{2}{\t^2}(  n(\phi) \cdot L \partial_A + 3  \partial_A n(\phi) \cdot L ) \sigmazero^+(\phi).
\ee
Using the relation $\izero^+_{\t A}  =  - 2  \izero^+_{A} $ one recovers the $\t \to +\infty$ identity in \eqref{delLitauAplus}.  The $\t \to -\infty$ limit can be obtained from the previous one  by doing the replacements $n^\mu \to \A_* n^\mu$ and $\sigmazero^+ \to \sigmazero^- $. 

\section{Matching infinities} \label{matchingapp}

In this appendix we review the asymptotic changes of coordinates \cite{cgw} that interpolate between Beig–Schmidt and Bondi coordinates near each pair of adjacent spacetime boundaries.
The analysis requires dealing with double limits: a first limit that takes us from the bulk spacetime into one of the $3d$ boundaries \eqref{5bdies}, and a second one that reaches the $2d$ boundaries \eqref{bdiesident}.  For instance, in  Bondi coordinates we take
\be \label{doublelimitbondigral}
|r |\to \infty, \quad |u| \to \infty, \quad \text{with } \quad |r|/|u| \to \infty,
\ee
where the last condition specifies the order of  limits. 

We will discuss separately  the  timelike and spatial infinity cases, with a focus on the pieces that are relevant for the derivation of soft theorems. 
In both cases, the starting point is the doubly-expanded  Bondi metric, obtained by considering the  large $u$ expansion of  \eqref{bondimetric},
\begin{multline} \label{largeuBondi}
ds^2_{\bondi}= (- 2 +\cdots) du dr + \Big(-1 + \frac{2 G  \Mzero}{r} + \cdots \Big) du^2 + \Big(r^2 q_{AB} +r \, \Czero_{AB} + \cdots \Big) d \O^A d \O^B\\
+ \left( D^B \Czero_{AB}+  \frac{4 G }{3 r} \big( \log |r|  \ov{\log r}{\N_A}+ \log |u| \ov{\log u}{\N_A} \big)+\cdots \right) d u  d \O^A ,
\end{multline}
where the dots stand for terms that are subleading under \eqref{doublelimitbondigral}.   We  denote by $\O^A$  the angular Bondi coordinates in order to distinguish them from the  Beig-Schmidt  angular coordinates $\phi^A$.

\subsection*{Timelike to null}

We start with  the asymptotic change of  coordinates between timelike and null infinities
\be \label{choctime}
(\t,\rho,\phi^A) \to (r,u,\O^A),
\ee
 within their overlapping region.  On the future/past timelike infinity side, this  region is characterized  by  
\be \label{doublelimittime}
\t \to \pm \infty, \quad \rho \to \infty, \quad \text{with } \quad \t/\rho \to \pm \infty,
\ee
where the last condition specifies the required order of limits.   Similarly, on the future/past null infinity side we  consider 
\be
r \to \pm \infty, \quad u \to \pm \infty, \quad \text{with } \quad r/u \to \infty.
\ee

To leading order, the change of coordinates \eqref{choctime}  can be identified by comparing the asymptotic Cartesian expressions \eqref{Xmuitotauxa} and  \eqref{Xmunullsec}. The large $\rho$ expansion of the unit timelike vector \eqref{Vxtime} reads
\be \label{Vlargerhomatch}
V^\mu(\rho,\phi)  \stackrel{\rho \to \infty}{=} \rho \, n^\mu(\phi) + \frac{1}{2 \rho} t^\mu + O(1/\rho^3),
\ee
where $t^\mu = (1,\vec{0})$. Substituting \eqref{Vlargerhomatch} into \eqref{Xmuitotauxa} leads to
\be \label{Xmutimedoublelim}
X^\mu = \tau  \rho \, n^\mu + \frac{\tau}{2 \rho} t^\mu + \cdots, 
\ee
where the dots are terms that are subleading under \eqref{doublelimittime}.  Comparing \eqref{Xmutimedoublelim} with \eqref{Xmunullsec} we learn that
\be \label{leadingchoctime}
r = \tau  \rho + \cdots, \quad u = \frac{\tau}{2 \rho} + \cdots , \quad \O^A = \phi^A + \cdots.
\ee

To specify the subleading terms in \eqref{leadingchoctime} one needs to set up a generic double asymptotic expansion for the change of coordinates  that maps the large-$u$ Bondi metric \eqref{largeuBondi} into the large-$\rho$ BS metric in radiative log frame. From the analysis of section \ref{timesec}, one finds the latter is given by\footnote{In addition to the fall-offs  $\sigma=O(\rho^{-3})$ and $i_{\rho A}=O(\rho^{-3})$, we are  using that  $\Phi=O(\rho)$ and $i_{\rho \rho}=O(\rho^{-6})$.}
\ba
g_{\t \rho} & = & g_{\t A} = 0 \\
g_{\t\t} &=& -1 - \frac{2 \sigmazero}{\t \rho^3} + \cdots \label{gttBStime} \\
g_{AB} &=& \t^2 \rho^2 q_{AB} + \cdots \\
g_{\rho A} &=& \t O(\rho^0) + \frac{\ln |\t|}{\rho^3} \ov{0}{i}_{\rho A} + \cdots \label{gtrhoBStime}\\
g_{\rho \rho} &=& \t^2 O(\rho^{-2})+ \t O(\rho^{-3}) +  \ln |\t| O(\rho^{-2}) + \cdots.
\ea

Consistency with \eqref{largeuBondi} requires 
\bal \label{asymchoctime}
r &= \tau  \rho + \t^0 O(\rho^{-1}) + \cdots \\
u &= \tau \rho  (\sqrt{1+\rho^{-2}}-1) + \t^0 O(\rho^{-3}) + \cdots \\
\O^A &= \phi^A + \t^{-1}O(\rho^{-3}) + \frac{\ln |\t|}{\t^2}\left( \frac{1}{\rho^4} \ov{\log}{\phi}^A + \cdots \right) + \cdots. 
\eal
where we have only displayed the coefficients that are needed for our analysis, see \cite{cgw} for a more complete treatment. The corresponding differentials are 
\bal \label{diffsasymchoctime}
d r &= d \tau  \rho + \tau d \rho + \cdots  \\
d u &= \frac{d\tau}{2 \rho} -\frac{\t d \rho}{2 \rho^2}  + \cdots \\
d \O^A &= d \phi^A + \cdots + \ln |\t| \ov{\log}{\phi}^A  \big(-2   \frac{d \t}{\t^3 \rho^4}-4 \frac{d \rho}{\t^2 \rho^5} \big) + \cdots 
\eal

Substituting  in \eqref{largeuBondi} and collecting coefficients of $d \t^2$ we recover \eqref{gttBStime} with
\be 
\sigmazero = - \frac{G}{4}  \Mzero,
\ee
which corresponds to Eq. \eqref{sigmamasstime} after including  the implicit labels.

Similarly, collecting $d \t d \phi^A$ and $d \rho \, d \phi^A$ terms one finds
\ba
g_{\t A} &= & \cdots + \frac{\ln |\t|}{\t  \rho^2}\left( \frac{G}{3 }  \left(\ov{\log r}{\N_A}+  \ov{\log u}{\N_A}  \right) -2 \ov{\log}{\phi}_A \right) + \cdots \label{gtauAchoctime} \\
g_{\rho A} &= & \cdots +  \frac{\ln |\t|}{\rho^3}\left( -\frac{G}{3 }  \left(\ov{\log r}{\N_A}+  \ov{\log u}{\N_A}  \right) -4 \ov{\log}{\phi}_A \right) + \cdots , \label{grhoAchoctime}
\ea
where, again, we only display terms that are needed for our analysis. The vanishing of \eqref{gtauAchoctime} determines $\ov{\log}{\phi}_A$, which upon substitution in \eqref{grhoAchoctime} leads to an expression consistent with \eqref{gtrhoBStime} with
\be 
\ov{0}{i}_{\rho A} =  -G \left(\ov{\log r}{\N_A}+  \ov{\log u}{\N_A}  \right).
\ee
Recalling \eqref{logangmomasp} and restoring labels, this leads to Eq. \eqref{irhoAangasptime}.

\subsection*{Spacelike to null}

We now present the map  between  asymptotic coordinates at spatial and null infinities
\be \label{chocspace}
(\rho,\t,\phi^A) \to (r,u,\O^A),
\ee
 within their common region
\be \label{doublelimitspace}
\rho \to \infty, \quad \t \to \pm \infty,  \quad \text{with } \quad \rho/\t \to \pm \infty,
\ee
and 
\be
r \to \pm \infty, \quad u \to \mp \infty, \quad \text{with } \quad r/u \to -\infty.
\ee

As before, to leading order the  change of coordinates  can be determined by comparing the  asymptotic Cartesian expressions \eqref{Xmuitotauxa} and  \eqref{Xmunullsec}. This leads to
\be \label{leadingruchocspace}
r =   \rho \, \t + \cdots, \quad u = -\frac{\rho}{2 \t} + \cdots ,
\ee
\be \label{leadingphichoc}
 \O^A = \begin{cases}
			\phi^A + \cdots, & \text{for} \quad \t \to +\infty\\
            \A \phi^A+ \cdots, & \text{for} \quad \t \to -\infty
            \end{cases}
\ee
where $\A \phi^A$ denotes the antipodal point of $\phi^A$.  Condition \eqref{leadingphichoc} ensures the future and past null directions are consistent with \eqref{limtaupinfVmu} and \eqref{limtauminfVmu} respectively. 
Including subleading terms, the asymptotic change of coordinates takes the form (we keep implicit for now the  antipodal map  \eqref{leadingphichoc} in the $\t \to -\infty$ case) 
\bal \label{asymchocspace}
r &=  \rho \,\tau \sqrt{1+\t^{-2}} + \rho^0 O(\t^{-1}) + \cdots \\
u &= - \rho \,\tau (\sqrt{1+\t^{-2}}-1) + \rho^0 O(\t^{-3}) + \cdots \\
\O^A &= \phi^A + \rho^{-1}O(\t^{-3}) + \frac{\ln \rho}{\rho^2}\left( \frac{1}{\t^4} \ov{\log}{\phi}^A + \cdots \right) + \cdots. 
\eal
with corresponding differentials given by
\bal \label{diffsasymchocspace}
d r &=  d \rho \, \t+ \rho\, d \tau    + \cdots  \\
d u &= -\frac{d\rho}{2 \tau} +\frac{\rho \, d \t}{2 \t^2}  + \cdots \\
d \O^A &= d \phi^A + \cdots + \ln \rho \ov{\log}{\phi}^A  \big(-2   \frac{d \rho}{\rho^3 \t^4}-4 \frac{d \t}{\rho^2 \t^5} \big)+ \cdots 
\eal

On the other hand, the BS metric coefficients \eqref{BSgaugespatial}, \eqref{gspatialab} in the limit \eqref{doublelimitspace} take the form
\ba
g_{\rho \t} & = & g_{\rho A} = 0 \\
g_{\rho\rho} &=& 1 + \frac{2 \sigmazero}{\rho \, \t^3} + \cdots \label{gttBStime} \\
g_{AB} &=& \rho^2 \t^2 q_{AB} + \cdots \\
g_{\t A} &=& \rho \,O(\t^0) + \frac{\ln \rho}{\t^3} \, \ov{0}{i}_{\t A} + \cdots \label{gtrhoBStime}\\
g_{\t \t} &=& \rho^2 O(\t^{-2})+ \rho  \,O(\t^{-3}) +  \ln \rho \, O(\t^{-2}) + \cdots.
\ea

As before, one can recover this expansion by applying the change of coordinates \eqref{asymchocspace} to the Bondi metric \eqref{largeuBondi}. We omit the steps as they are essentially the same as those presented in the timelike case. The resulting matching conditions are those presented in Eqs. \eqref{sigmamassspip}, \eqref{sigmamassspim}, \eqref{itAitoNAspip}, \eqref{itAitoNAspim}.

\section{Sphere derivatives of  soft factors} \label{2didsapp}

In this section we discuss the identities used for the evaluation of Eqs.  \eqref{leadingthmrew} and \eqref{logsoftthmrewproof}.

The identities are most easily established in a  2d  frame in which the celestial metric is flat (see e.g. appendix A of \cite{Kapec:2017gsg}). This can be achieved by choosing  the null vector $n^\mu$ to be
\be 
n^\mu = \frac{1}{\sqrt{2}}\left( 1+ |z|^2, z+ \zb, -i (z- \zb), 1- |z|^2 \right),
\ee
where $(z,\zb)$ are stereographic coordinates on the celestial sphere. 

We take
\be 
k^\mu = \frac{1}{\sqrt{2}} \left(1,0,0,-1 \right),
\ee
to be a complementary null direction such that $k \cdot n =-1$.  Together with the polarization vectors $\partial_z n^\mu$ and $\partial_{\zb} n^\mu$, these give a (complex) basis of  four null vectors. 
The  resolution of the identity in this basis can be written as
\be
\delta^\mu_\nu  = \ov{+}{\delta}^{ \mu}_\nu+ \ov{-}{\delta}^{ \mu}_\nu
\ee
where
\ba
\ov{+}{\delta}^{ \mu}_\nu & = & \partial^z n^\mu   \partial_z n_\nu  - k^\mu n_\nu ,\\
\ov{-}{\delta}^{ \mu}_\nu & = & \partial^{\zb} n^\mu   \partial_{\zb} n_\nu  - n^\mu k_\nu ,
\ea
are projections on the null planes $(k^\mu,\partial^{z} n^\mu)$ and $(n^\mu,\partial^{\zb} n^\mu)$,  and $\partial^z  \equiv \partial_{\zb}$.   

All the identities we shall need are a consequence of the following ``master'' identity that may be checked by a short computation:
\be \label{partial2zplus}
\partial_z^2 \left(     \frac{\partial^z n^\mu  \partial^z n^\nu    }{p \cdot n} \right) = \frac{2 \ov{+}{p}^{ \mu} \ov{+}{p}^{ \nu}}{(p \cdot n)^3} ,
\ee
where $p^\mu$ is a \emph{massive} momentum and $\ov{\pm}{p}^{\mu}=\ov{\pm}{\delta}^{ \mu}_\nu p^\nu$.   For later use,  we note that
\be \label{idspluspmin}
p^\mu = \ov{+}{p}^{\mu}+ \ov{-}{p}^{\mu}, \quad  \ov{\pm}{p}\cdot \ov{\pm}{p} =0, \quad  \ov{+}{p}\cdot \ov{-}{p} =-m^2/2.
\ee

Consider first the contraction of  \eqref{partial2zplus} with $p_\mu p_\nu$. Using  \eqref{idspluspmin} one finds 
\be \label{partial2zS0}
\partial_z^2 \left(     \frac{\partial^z n \cdot p \, \partial^z n \cdot p  }{p \cdot n} \right) =  \frac{m^4}{2}\frac{1}{(p \cdot n)^3}.
\ee
  The massless limit of this relation has been discussed in  \cite{clmassive,mcgreen} and gives  \cite{stromST}
\be
\partial_z^2 \left(     \frac{\partial^z n \cdot p \, \partial^z n \cdot p  }{p \cdot n} \right) \stackrel{m\to 0}{=} - 2 \pi E \delta(\nh,\ph), 
\ee
where $\ph = \vec{p}/E$, and  $E$  positive (negative) for outgoing (incoming) momentum.

For the log soft factor we have an extra derivative and an extra factor of the null vector $n^\rho$. Since  $\partial_z^2 n^\rho=0$ we have that
\be \label{partialz3nf}
\partial_z^3 (n^\rho f) = \left(n^\rho \partial_z + 3 \partial_z n^\rho\right) \partial_z^2 f
\ee
for any function $f$. In particular,
\ba 
\partial^3_z \left(n^\rho \frac{\partial^z n^\mu  \partial^z n^\nu }{p \cdot n} \right)  &=&  \left(n^\rho \partial_z + 3 \partial_z n^\rho\right)  \frac{2\ov{+}{p}^{ \mu} \ov{+}{p}^{ \nu}}{(p \cdot n)^3} \label{partial3zplus} \\
&=&    \frac{6   \ov{+}{p}^{ \mu} \ov{+}{p}^{ \nu}  \left(\partial_z n^\rho p \cdot n - n^\rho \, p \cdot \partial_z n\right)}{(p \cdot n)^4},\label{partial3zplusexpanded} 
\ea
where in the first equality  we used  Eq. \eqref{partial2zplus} and in the second equality we evaluated the derivative in \eqref{partial3zplus}.

To deal with the second line of \eqref{hlnasymfield},   we contract \eqref{partial3zplus} with $p_\mu p_\nu b_\rho$ with $b_\rho$  a constant vector. This results in
\ba
\partial^3_z \left(n \cdot b \frac{\partial^z n \cdot p \,  \partial^z n \cdot p }{p \cdot n} \right) &= &\frac{m^4}{2} \left(n \cdot b \partial_z + 3 \partial_z n \cdot b\right) \frac{1}{(p \cdot n)^3}  \label{d3zdrag}\\ 
&= & \frac{3}{2}  m^4 \frac{  \partial_z n^{\mu} n^{\nu} b_{[\mu} p_{\nu]}}{(p \cdot n)^4}. \label{1stcontractionp3z}
\ea
This relation  with $b_\mu = P_\mu$, explains the first term in the second line of \eqref{logsoftthmrewproof} (the second term in that line follows simply from \eqref{partialz3nf}).

To handle  the ``angular momentum" contribution in \eqref{hlnasymfield}, we contract \eqref{partial3zplus} with $p_{\mu}  J_{\nu \rho}$ with
\be \label{auxiliaryJ}
J_{\nu \rho}= b_\nu p_\rho -b_\rho p_\nu  ,
\ee
for a constant vector $b_\mu$.  It easy to see from \eqref{partial3zplusexpanded} that only the second term in \eqref{auxiliaryJ} contributes in the contraction, thus reducing the result to (minus) that  in \eqref{1stcontractionp3z},
\be \label{2ndcontractionp3z}
\partial^3_z \left( \frac{\partial^z n \cdot p \,  \partial^z n^\nu n^\rho J_{\nu \rho} }{p \cdot n} \right)  =  - \frac{3}{2}  m^4 \frac{  \partial_z n^{\mu} n^{\nu} b_{[\mu} p_{\nu]}}{(p \cdot n)^4}.
\ee
This relation, with $b_\mu=c^{\rad \pm}_\mu$, leads to the first two terms in \eqref{logsoftthmrewproof}.


\begin{thebibliography}{99}


\bibitem{penrose}
R.~Penrose,
``Asymptotic properties of fields and space-times,''
Phys. Rev. Lett. \textbf{10}, 66-68 (1963)


  
 \bibitem{aaprl} 
  A.~Ashtekar,
  ``Asymptotic Quantization of the Gravitational Field,''
  Phys.\ Rev.\ Lett.\  {\bf 46}, 573 (1981)

\bibitem{AS} 
  A.~Ashtekar and M.~Streubel,
  ``Symplectic Geometry of Radiative Modes and Conserved Quantities at Null Infinity,''
  Proc.\ Roy.\ Soc.\ Lond.\ A {\bf 376}, 585 (1981)

\bibitem{aajmp} 
  A.~Ashtekar,
  ``Radiative Degrees of Freedom of the Gravitational Field in Exact General Relativity,''
  J.\ Math.\ Phys.\  {\bf 22}, 2885 (1981)



\bibitem{stromgravscatt}
A.~Strominger,
``On BMS Invariance of Gravitational Scattering,''
JHEP \textbf{07}, 152 (2014)



\bibitem{stromlectures}
A.~Strominger,
``Lectures on the Infrared Structure of Gravity and Gauge Theory,''
Princeton University Press, 2018,


  \bibitem{laddhasen1}
A.~Laddha and A.~Sen,
``Logarithmic Terms in the Soft Expansion in Four Dimensions,''
JHEP \textbf{10}, 056 (2018)


\bibitem{sahoosen}
B.~Sahoo and A.~Sen,
``Classical and Quantum Results on Logarithmic Terms in the Soft Theorem in Four Dimensions,''
JHEP \textbf{02}, 086 (2019)


\bibitem{proofdeq4}
A.~P.~Saha, B.~Sahoo and A.~Sen,
``Proof of the classical soft graviton theorem in $D$ = 4,''
JHEP \textbf{06}, 153 (2020)


  
 \bibitem{senreview}
A.~Sen,
``Gravitational Wave Tails from Soft Theorem: A Short Review,''
[arXiv:2408.08851 [hep-th]]. 



\bibitem{logwcl}
M.~Campiglia and A.~Laddha,
``Loop Corrected Soft Photon Theorem as a Ward Identity,''
JHEP \textbf{10}, 287 (2019)


\bibitem{sayalicons}
S.~Atul Bhatkar,
``New asymptotic conservation laws forelectromagnetism,''
JHEP \textbf{02}, 082 (2021)


\bibitem{sayaliqedgrav}
S.~Atul Bhatkar,
``Ward identity for loop level soft photon theorem for massless QED coupled to gravity,''
JHEP \textbf{10}, 110 (2020)



\bibitem{comperelogem}
G.~Comp{\`e}re, D.~Fontaine and K.~Nguyen,
``Electromagnetic multipole expansions and the logarithmic soft photon theorem,''
SciPost Phys. Core \textbf{8}, 066 (2025)

\bibitem{fuentelogscalar}
O.~Fuentealba and M.~Henneaux,
``Logarithmic matching between past infinity and future infinity: The massless scalar field in Minkowski space,''
JHEP \textbf{03}, 081 (2025)

\bibitem{bricenolog}
M.~Brice{\~n}o, H.~A.~Gonz{\'a}lez, M.~Henneaux and A.~P{\'e}rez,
``Matching conditions at null infinity in the presence of logarithms: the role of advanced and retarded radiation,''
[arXiv:2510.21072 [hep-th]].


\bibitem{Duary:2025siq}
S.~Duary and P.~Ray,
``Waveform, memory and classical soft scalar theorems,''
Nucl. Phys. B \textbf{1018}, 117044 (2025)

\bibitem{cgw}
G.~Comp\`ere, S.~E.~Gralla and H.~Wei,
``An asymptotic framework for gravitational scattering,''
Class. Quant. Grav. \textbf{40}, no.20, 205018 (2023)

\bibitem{hansen}
A.~Ashtekar and R.~O.~Hansen,
``A unified treatment of null and spatial infinity in general relativity. I - Universal structure, asymptotic symmetries, and conserved quantities at spatial infinity,''
J. Math. Phys. \textbf{19}, 1542-1566 (1978)

\bibitem{persides}
S.~Persides,
``A unified formulation of timelike, null and spatial infinity,''
J. Math. Phys. \textbf{23}, no.2, 289-292 (1982)


\bibitem{compererobert}
G.~Comp{\`e}re and S.~Robert,
``A proof of conservation laws in gravitational scattering: tails and breaking of peeling,''
[arXiv:2603.08705 [hep-th]].


\bibitem{Kim:2023qbl}
S.~Kim, P.~Kraus, R.~Monten and R.~M.~Myers,
``S-matrix path integral approach to symmetries and soft theorems,''
JHEP \textbf{10}, 036 (2023)


\bibitem{Jain:2023fxc}
D.~Jain, S.~Kundu, S.~Minwalla, O.~Parrikar, S.~G.~Prabhu and P.~Shrivastava,
``The S-matrix and boundary correlators in flat space,''
[arXiv:2311.03443 [hep-th]].

\bibitem{Kraus:2024gso}
P.~Kraus and R.~M.~Myers,
``Carrollian partition functions and the flat limit of AdS,''
JHEP \textbf{01}, 183 (2025)

\bibitem{Ammon:2025avo}
M.~Ammon, F.~Capone and C.~Sieling,
``Flat Holography {\&} Holographic Renormalization: Scalar Field,''
[arXiv:2512.14818 [hep-th]].




\bibitem{raju}
A.~Laddha, S.~G.~Prabhu, S.~Raju and P.~Shrivastava,
``The Holographic Nature of Null Infinity,''
SciPost Phys. \textbf{10}, no.2, 041 (2021)
[arXiv:2002.02448 [hep-th]].


\bibitem{Donnay:2023mrd}
L.~Donnay,
``Celestial holography: An asymptotic symmetry perspective,''
Phys. Rept. \textbf{1073}, 1-41 (2024)


\bibitem{Bagchi:2023cen}
A.~Bagchi, P.~Dhivakar and S.~Dutta,
``Holography in flat spacetimes: the case for Carroll,''
JHEP \textbf{08}, 144 (2024)



\bibitem{Bergmann:1961zz}
P.~G.~Bergmann,
``'Gauge-Invariant' Variables in General Relativity,''
Phys. Rev. \textbf{124}, 274-278 (1961)


\bibitem{aalog}
A. Ashtekar, ``Logarithmic ambiguities in the description of spatial infinity", Found. Phys. \textbf{15}, 419–431 (1985)





\bibitem{gianni1}
G.~Boschetti and M.~Campiglia,
``Log translation invariance of log soft gravitational radiation,''
JHEP \textbf{10}, 105 (2025)





\bibitem{blanchetnull}
L.~Blanchet,
``Radiative gravitational fields in general relativity. 2. Asymptotic behaviour at future null infinity,''
Proc. Roy. Soc. Lond. A \textbf{409}, 383-399 (1987)


\bibitem{tn}
A. Ashtekar and R. Penrose, “Mass Positivity from Focussing and the Structure of io ”,
Twistor Newsletter, 31 (1991)



 \bibitem{weinberg}
S.~Weinberg,
``Infrared photons and gravitons,''
Phys. Rev. \textbf{140}, B516-B524 (1965)
 
 

\bibitem{thorne}
V.~B.~Braginsky and K.~S.~Thorne,
``Gravitational-wave bursts with memory and experimental prospects,''
Nature \textbf{327}, 123-125 (1987)


 
\bibitem{zhibo}
A.~Strominger and A.~Zhiboedov,
``Gravitational Memory, BMS Supertranslations and Soft Theorems,''
JHEP \textbf{01}, 086 (2016)
  
 
\bibitem{stromingercachazo}
F.~Cachazo and A.~Strominger,
``Evidence for a New Soft Graviton Theorem,''
[arXiv:1404.4091 [hep-th]].


\bibitem{laddhasen}
A.~Laddha and A.~Sen,
``Observational Signature of the Logarithmic Terms in the Soft Graviton Theorem,''
Phys. Rev. D \textbf{100}, no.2, 024009 (2019)
[arXiv:1806.01872 [hep-th]].


\bibitem{rewritten}
B.~Sahoo and A.~Sen,
``Classical soft graviton theorem rewritten,''
JHEP \textbf{01}, 077 (2022)



\bibitem{stromST} 
  T.~He, V.~Lysov, P.~Mitra and A.~Strominger,
  ``BMS supertranslations and Weinberg’s soft graviton theorem,''
  JHEP {\bf 1505}, 151 (2015)



\bibitem{BS}
R.~Beig and B.~G.~Schmidt,
``Einstein's equations near spatial infinity,''
Commun. Math. Phys. \textbf{87}, no.1, 65-80 (1982)




\bibitem{dehouck}
G.~Compere and F.~Dehouck,
``Relaxing the Parity Conditions of Asymptotically Flat Gravity,''
Class. Quant. Grav. \textbf{28}, 245016 (2011)
[erratum: Class. Quant. Grav. \textbf{30}, 039501 (2013)]


\bibitem{phigoldstone}
K.~Nguyen and J.~Salzer,
``Celestial IR divergences and the effective action of supertranslation modes,''
JHEP \textbf{09}, 144 (2021)


\bibitem{mcgreen}
M.~Campiglia,
``Null to time-like infinity Green{\textquoteright}s functions for asymptotic symmetries in Minkowski spacetime,''
JHEP \textbf{11}, 160 (2015)


\bibitem{Compere:2019gft}
G.~Comp{\`e}re, R.~Oliveri and A.~Seraj,
``The Poincar{\'e} and BMS flux-balance laws with application to binary systems,''
JHEP \textbf{10}, 116 (2020)



\bibitem{clmassive}
M.~Campiglia and A.~Laddha,
``Asymptotic symmetries of gravity and soft theorems for massive particles,''
JHEP \textbf{12}, 094 (2015)


\bibitem{chipum2}
S.~Choi, A.~Laddha and A.~Puhm,
``The classical super-rotation infrared triangle. Classical logarithmic soft theorem as conservation law in gravity,''
JHEP \textbf{04}, 138 (2025)


\bibitem{romano}
A.~Ashtekar and J.~D.~Romano,
``Spatial infinity as a boundary of space-time,''
Class. Quant. Grav. \textbf{9}, 1069-1100 (1992)


\bibitem{mmvirmani}
R.~B.~Mann, D.~Marolf and A.~Virmani,
``Covariant Counterterms and Conserved Charges in Asymptotically Flat Spacetimes,''
Class. Quant. Grav. \textbf{23}, 6357-6378 (2006)


\bibitem{magnonenergymom}
A.~Ashtekar and A.~Magnon-Ashtekar,
``Energy-Momentum in General Relativity,''
Phys. Rev. Lett. \textbf{43}, no.3, 181 (1979)





\bibitem{cedric}
C.~Troessaert,
``The BMS4 algebra at spatial infinity,''
Class. Quant. Grav. \textbf{35}, no.7, 074003 (2018)


\bibitem{prabhu}
K.~Prabhu,
``Conservation of asymptotic charges from past to future null infinity: Supermomentum in general relativity,''
JHEP \textbf{03}, 148 (2019)


\bibitem{capone}
F.~Capone, K.~Nguyen and E.~Parisini,
``Charge and antipodal matching across spatial infinity,''
SciPost Phys. \textbf{14}, no.2, 014 (2023)
[arXiv:2204.06571 [hep-th]].


\bibitem{herber}
M.~Herberthson and M.~Ludvigsen,
``A relationship between future and past null infinity,''
Gen. Rel. Grav. \textbf{24}, no.11, 1185-1193 (1992)




\bibitem{bondi} 
  H.~Bondi, M.~G.~J.~van der Burg and A.~W.~K.~Metzner,
  ``Gravitational waves in general relativity. 7. Waves from axisymmetric isolated systems,''
  Proc.\ Roy.\ Soc.\ Lond.\ A {\bf 269}, 21 (1962).

\bibitem{sachs} 
  R.~K.~Sachs,
  ``Gravitational waves in general relativity. 8. Waves in asymptotically flat space-times,''
  Proc.\ Roy.\ Soc.\ Lond.\ A {\bf 270}, 103 (1962).




\bibitem{winicour} 
J. Winicour, ``Logarithmic asymptotic flatness'',
Found. Phys. {\bf 15}, 605–616 (1985).


\bibitem{damour}
T. Damour, ``Analytical calculations of gravitational radiation'', in Fourth Marcel Grossmann
Meeting on General Relativity, pp. 365–392, Jan., 1986.


\bibitem{geillerpeeling}
M.~Geiller, A.~Laddha and C.~Zwikel,
``Symmetries of the gravitational scattering in the absence of peeling,''
JHEP \textbf{12}, 081 (2024)


\bibitem{BT} 
  G.~Barnich and C.~Troessaert,
  ``Aspects of the BMS/CFT correspondence,''
  JHEP {\bf 1005}, 062 (2010)


\bibitem{gianni3} G.~Boschetti and M.~Campiglia, ``Peeling-violating coefficients in gravitational scattering'', \emph{in progress}.




\bibitem{compnich}
G.~Comp{\`e}re and D.~A.~Nichols,
``Classical and Quantized General-Relativistic Angular Momentum,''
[arXiv:2103.17103 [gr-qc]].


\bibitem{HPS}
S.~W.~Hawking, M.~J.~Perry and A.~Strominger,
``Superrotation Charge and Supertranslation Hair on Black Holes,''
JHEP \textbf{05}, 161 (2017)



\bibitem{stromvirasoro}
D.~Kapec, V.~Lysov, S.~Pasterski and A.~Strominger,
``Semiclassical Virasoro symmetry of the quantum gravity $ \mathcal{S}$-matrix,''
JHEP \textbf{08}, 058 (2014)


\bibitem{relaxed}
M.~Campiglia and A.~Laddha,
``Asymptotic symmetries and subleading soft graviton theorem,''
Phys. Rev. D \textbf{90}, no.12, 124028 (2014)


\bibitem{eyhe}
M.~Campiglia and R.~Eyheralde,
``Asymptotic $U(1)$ charges at spatial infinity,''
JHEP \textbf{11}, 168 (2017)


\bibitem{coito}
M.~Campiglia, L.~Coito and S.~Mizera,
``Can scalars have asymptotic symmetries?,''
Phys. Rev. D \textbf{97}, no.4, 046002 (2018)


\bibitem{Briceno:2025ivl}
M.~Brice{\~n}o, H.~A.~Gonz{\'a}lez and A.~P{\'e}rez,
``Scalar subleading soft theorems from an infinite tower of charges,''
[arXiv:2504.08612 [hep-th]].


\bibitem{comprobds3}
G.~Comp{\`e}re and S.~Robert,
``Scalar, vector and tensor fields on $dS_3$ with arbitrary sources: harmonic analysis and antipodal maps,''
[arXiv:2512.15578 [hep-th]].




\bibitem{henntrem}
M.~Henneaux and C.~Troessaert,
``Asymptotic symmetries of electromagnetism at spatial infinity,''
JHEP \textbf{05}, 137 (2018)


\bibitem{henntrscalar}
M.~Henneaux and C.~Troessaert,
``Asymptotic structure of a massless scalar field and its dual two-form field at spatial infinity,''
JHEP \textbf{05}, 147 (2019)



\bibitem{deBoer:2003vf}
J.~de Boer and S.~N.~Solodukhin,
``A Holographic reduction of Minkowski space-time,''
Nucl. Phys. B \textbf{665}, 545-593 (2003)


\bibitem{Pasterski:2017kqt}
S.~Pasterski and S.~H.~Shao,
``Conformal basis for flat space amplitudes,''
Phys. Rev. D \textbf{96}, no.6, 065022 (2017)



\bibitem{Donnay:2018neh}
L.~Donnay, A.~Puhm and A.~Strominger,
``Conformally Soft Photons and Gravitons,''
JHEP \textbf{01}, 184 (2019)



\bibitem{Pasterski:2020pdk}
S.~Pasterski and A.~Puhm,
``Shifting spin on the celestial sphere,''
Phys. Rev. D \textbf{104}, no.8, 086020 (2021)



\bibitem{aview}
M.~Campiglia and A.~Laddha,
``Asymptotic charges in massless QED revisited: A view from Spatial Infinity,''
JHEP \textbf{05}, 207 (2019)


\bibitem{gelfandshilov}
M. Gel'fand and G. E. Shilov,
``Generalized functions'', vol. 1 and 3. Academic Press, New York and London, 1977.


\bibitem{Kapec:2017gsg}
D.~Kapec and P.~Mitra,
``A $d$-Dimensional Stress Tensor for Mink$_{d+2}$ Gravity,''
JHEP \textbf{05}, 186 (2018)

\end{thebibliography}
\end{document}